\documentclass[a4paper,11pt]{article}
\pdfoutput=1 

\usepackage{jcappub} 

\usepackage[T1]{fontenc} 
\usepackage{graphicx}
\usepackage{epstopdf}
\usepackage{upgreek}

\usepackage{makecell,multirow,diagbox}

\title{\boldmath Inversion of stellar spectral radiative properties based on multiple star catalogues}

\author[a]{Chuanxin Zhang,}
\author[a,*]{Yuan Yuan,\note{*Corresponding author.}}
\author[a]{Zhaoyang Yu,}
\author[b,*]{Fuqiang Wang}
\author[a]{and Heping Tan}

\affiliation[a]{Key Laboratory of Aerospace Thermophysics, Ministry of Industry and Information Technology, School of Energy Science and Engineering, Harbin Institute of Technology,\\92 West Dazhi Street, Harbin 150001, China}
\affiliation[b]{School of Automobile Engineering, Harbin Institute of Technology at Weihai,\\2 West Wenhua Road, Weihai 264209, China}

\emailAdd{yuanyuan83@hit.edu.cn}
\emailAdd{wangfuqiang@hitwh.edu.cn}

\abstract{The spectral flux density of stars can indicate their atmospheric physical properties. A detector can obtain any band flux density at the design stage. However, the band flux density is confirmed and fixed in the process of operation because of the restriction of filters. Other band flux densities cannot be obtained through the same detector. In this study, a computational model of stellar spectral flux density is established based on basic physical parameters which are effective temperature and angular parameter. The stochastic particle swarm optimization algorithm is adopted to address this issue with appropriately chosen values of the algorithm parameters. Four star catalogues are studied and consist of the Large Sky Area Multi-Object Fibre Spectroscopic Telescope (LAMOST), Wide-field Infrared Survey Explorer (WISE), Midcourse Space Experiment (MSX), and Two Micron All Sky Survey (2MASS). The given flux densities from catalogues are input parameters. Stellar effective temperatures and angular parameters are inverted using the given flux densities according to SPSO algorithm. Then the flux density is calculated according to Planck's law on the basis of stellar effective temperatures and angular parameters. The calculated flux density is compared with the given value from catalogues. It is found that the inversion results are in good agreement for all bands of the MSX and 2MASS catalogues, whereas they do not agree well in some bands of the LAMOST and WISE catalogues. Based on the results, data from the MSX and 2MASS catalogues can be used to calculate the spectral flux density at different wavelengths of given wavelength ranges. The stellar flux density is obtained and can provide data support and an effective reference for detection and recognition of stars.}

\keywords{radiative transfer, star catalogs, data analysis, detectors}

\begin{document}
\maketitle
\flushbottom

\section{Introduction}
\label{sec:intro}

The magnitude of stellar radiation energy is related to the surface and atmosphere of a star, and the measurement of flux density is the basis for determining its atmospheric physical properties ~\cite{a,b,c}. This relationship between stellar spectral flux density and the surface and physical properties is used to calculate the albedo, size, element abundance, temperature profile, and even interior conditions of stars ~\cite{d,e,f,g,h,i,j,k}. The emitted energy of a star is a function of wavelength or frequency, but it is difficult to measure the complete spectral energy distribution with a single detector. Only the flux density at a few finite wavelengths can be obtained by existing detectors primarily for the following two reasons. First, the radiation energy at different frequencies needs to be measured by different instruments, so the relative calibration can be problematic. Secondly, the radiative penetration depths through the Earth's atmosphere are different due to absorption and scattering ~\cite{l,m,n}. Spectral energy distribution data in the wide band range can be used to infer the properties of emitting source and the influence of the interstellar medium along the line of sight ~\cite{o,p}. The spectral flux density as a basic stellar apparent parameter has an important role in studying stellar physical properties ~\cite{q}. Only one fixed band detection is not sufficient for studying the physical properties of stellar atmospheres, confirming the composition of stellar chemical elements, or target detection and identification ~\cite{r,s,t}. A calculation model of stellar flux should be established to obtain the flux density, and the flux density at different wavelengths needs to be studied.

Some researchers have calculated the wavelength band flux density through the infrared radiation flux method (IRFM). Alonso et al. ~\cite{u} proposed a semi-empirical method to derive the absolute flux calibration in near-infrared bands. The method consists of the application of the IRFM to a selected sample of stars with accurate direct measurements of their angular diameters. Blackwell et al. ~\cite{v,w} discussed the relationships between stellar integral flux and the photometric indices B, V, I, and K, and derived formulae based on the database of measured integrated fluxes.

In addition, researchers usually establish a template matching method to calculate the flux density in each band based on the basic stellar parameters, including the effective temperature, element abundance, and detection distance. Ludwig et al. ~\cite{x} calculated the effective flux limited by using a one-dimensional radiation convection model and analysed the effective flux of stars with planets in the habitable zone. In these cases, stars with larger errors in effective flux, were the result of larger errors in the band radiation flux, affecting detection and identification. Therefore, it is necessary to study the spectral flux density of different wavelengths in given bands.

Based on the stellar spectral radiation properties, we constructed a calculation model of band flux density. In this paper, stellar flux density at different wavelengths of given bands is calculated using stochastic particle swarm optimization (SPSO) algorithm. The results of the SPSO calculations are used to optimize the band combinations. Four star catalogues (LAMOST, WISE, MSX, and 2MASS) are studied. The paper is organized as follows. In section~\ref{sec:method} we describe the stellar flux density model and SPSO algorithm. The results of stellar flux density are presented in section~\ref{sec:results}. In section~\ref{sec:conclusions} we draw our conclusions.

\section{Method and model}
\label{sec:method}
\begin{figure}[!htbp]
\setlength{\abovecaptionskip}{0.cm}
\setlength{\belowcaptionskip}{-0.cm}
   \centering
   \includegraphics[width=11.0cm, height=2.2cm, angle=0]{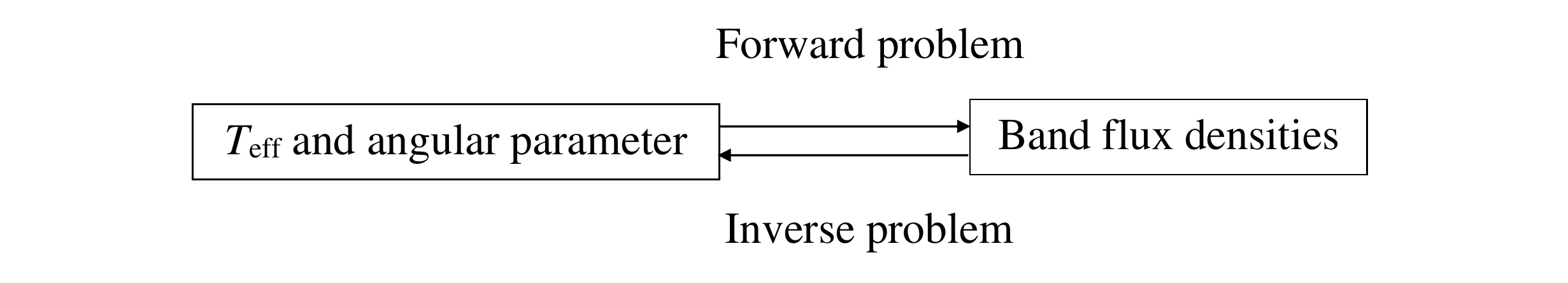}
   \caption{Forward and inverse problems of stellar flux densities.}
   \label{fig:1}
   \end{figure}

The calculation model of band flux density is established according to the stellar properties, which are $T_{eff}$ and angular parameter. Four star catalogues (LAMOST, WISE, MSX, 2MASS) are introduced and characteristic data of the detection band and flux density in these catalogues are extracted. These data provide input parameters for the SPSO algorithm to derive effective temperatures and angular parameters. Then the flux density is calculated by the above model. This relationship is shown in figure~\ref{fig:1}.

Physical model established in section~\ref{sec:stellar} is a forward problem, and the SPSO algorithm in section~\ref{sec:spso} is introduced to solve the inverse problem algorithm. The forward problem takes $T_{eff}$ and angular parameter as an input and gives flux densities. The inverse problem considers flux densities as an input and $T_{eff}$ and angular parameter as an output.

\subsection{Stellar flux density model}
\label{sec:stellar}
Stars have similar radiative properties to blackbodies. However, significant absorption or emission lines exist at some wavelengths because of the presence of gases in the stellar atmosphere with different temperatures, pressures, and densities ~\cite{y}. The stellar surface temperature can be calculated by referring to the blackbody radiative formula if the stellar atmosphere is in thermal equilibrium. To be more specific, there are differences of blackbody radiation in some bands. This deviation is defined as grey characteristic. The assumption that the star has a similar spectral emissivity and grey characteristic is practical only if these parameters satisfy the requirement for the accuracy of the probing results.

Planck's law describes the variation of blackbody spectral radiative power with wavelength, as shown in eq. \eqref{eq:2.1}. ${E_b}_{\left( {{\lambda _1} - {\lambda _2}} \right)}$ represents the blackbody spectral emissive power from $\lambda_1$ to $\lambda_2$ ( Wm$^{-2}$ ), $\lambda$ is the wavelength (m), \emph{T} is the thermodynamic temperature of the blackbody (K), \emph{c}$_1$ is the first radiation constant ($\emph{c}_1$ = 3.7419$\times$10$^{-16}$ Wm$^2$), and $\emph{c}_2$ is the second radiation constant ($\emph{c}_2$ = 1.4388419$\times$10$^{-2}$ mK).

The band emissivity is denoted by $\varepsilon$, and the radiative power is described by eq. \eqref{eq:2.2}. The radiative intensity is expressed as eq. \eqref{eq:2.3}. Radius in the stellar effective temperature calculation is \emph{r}, and the distance between the star surface and receiver surface is \emph{R}. The solid angle from the star surface to a detector is described by eq. \eqref{eq:2.4}. Radiative power received by a detector is expressed as eq. \eqref{eq:2.5}.

\begin{equation}
\label{eq:2.1}
{E_b}_{\left( {{\lambda _1} - {\lambda _2}} \right)} = \int_{{\lambda _1}}^{{\lambda _2}} {\frac{{{c_1}{\lambda ^{ - 5}}}}{{\exp [{c_2}/(\lambda T)] - 1}}} {\rm{d}}\lambda   
\end{equation}

\begin{equation}
\label{eq:2.2}
{E_{{\lambda _1} - {\lambda _2}}}{\rm{ = }}{\varepsilon _{{\lambda _1} - {\lambda _2}}} \cdot {E_b}_{\left( {{\lambda _1} - {\lambda _2}} \right)}  
\end{equation}

\begin{equation}
\label{eq:2.3}
{I_{{\lambda _1} - {\lambda _2}}}{\rm{ = }}{{{E_{{\lambda _1} - {\lambda _2}}}} \mathord{\left/
 {\vphantom {{{E_{{\lambda _1} - {\lambda _2}}}} \pi }} \right.
 \kern-\nulldelimiterspace} \pi }                                                                                                 
\end{equation}

\begin{equation}
\label{eq:2.4}
{\rm{d}}\Omega  = {{\pi {r^2}} \mathord{\left/
 {\vphantom {{\pi {r^2}} {{R^2}}}} \right.
 \kern-\nulldelimiterspace} {{R^2}}}                                                                                       
\end{equation}

\begin{equation}
\label{eq:2.5}
{E_{p\left( {{\lambda _1} - {\lambda _2}} \right)}} = {\rm{d}}\Omega  \cdot {I_{{\lambda _1} - {\lambda _2}}}                 
\end{equation}

The radiative power received by a detector can be written as eq. \eqref{eq:2.6} according to eq. \eqref{eq:2.1} to eq. \eqref{eq:2.5}. Thus, the angular parameter from Earth's surface to the star is defined as eq. \eqref{eq:2.7}. Eq. \eqref{eq:2.6} can be written as eq. \eqref{eq:2.8}. The radiative power received by a detector on Earth is determined by eq. \eqref{eq:2.8}. That is to say, the radiative power received by a detector is calculated using the stellar effective temperature $T_{eff}$ and angular parameter $\emph{x}_i$, as shown in eq. \eqref{eq:2.9}.

\begin{equation}
\label{eq:2.6}
{E_{p\left( {{\lambda _1} - {\lambda _2}} \right)}} = \frac{{{r^2}}}{{{R^2}}} \cdot {\varepsilon _{{\lambda _1} - {\lambda _2}}} \cdot \int_{{\lambda _1}}^{{\lambda _2}} {\frac{{{c_1}{\lambda ^{ - 5}}}}{{\exp [{c_2}/(\lambda T)] - 1}}} {\rm{d}}\lambda                          
\end{equation}

\begin{equation}
\label{eq:2.7}
{\xi _{{\lambda _1} - {\lambda _2}}} = \frac{{{r^2}}}{{{R^2}}} \cdot {\varepsilon _{{\lambda _1} - {\lambda _2}}}                     
\end{equation}

\begin{equation}
\label{eq:2.8}
{E_{p\left( {{\lambda _1} - {\lambda _2}} \right)}} = {\xi _{{\lambda _1} - {\lambda _2}}} \cdot \int_{{\lambda _1}}^{{\lambda _2}} {\frac{{{c_1}{\lambda ^{ - 5}}}}{{\exp [{c_2}/(\lambda T)] - 1}}} {\rm{d}}\lambda                      
\end{equation}

\begin{equation}
\label{eq:2.9}
{E_{p\left( {{\lambda _1} - {\lambda _2}} \right)}} = f\left( {T,\xi } \right)                      
\end{equation}

\subsection{Star catalogues}

The observations are recorded with specific filters and therefore only certain wavelength regions can be observed, whereas the light from the other bands is not recorded. To address this problem, if the stellar effective temperature and angular parameter can be acquired by the inversion method through the known flux density, the other band data can be calculated through the previous calculation model. Therefore, the physical model established in section~\ref{sec:stellar} is a forward problem, and the SPSO algorithm is introduced as an inverse problem algorithm to study the above problem. SPSO is used first to invert $T_{eff}$ and angular parameter from the available flux in different bands. And then using the Planck's law from eq. \eqref{eq:2.9}, the whole flux distribution in all the bands can be derived.

According to this idea, the forward problem process is described as follows. The radiative power is obtained from the known stellar effective temperature and angular parameter, then, the stellar average flux density is calculated. $T_{eff}$ and angular parameters are the known variables in the forward problem. The stellar effective temperature and angular parameter are obtained through the inversion method. The driving source of the inverse problem is the average flux density of a given band. In this study, the driven data of the inverse problem are derived from the star catalogues.

The observation of stars is divided into ground and space observation. Ground observation includes LAMOST ~\cite{z}, WISE ~\cite{a1}, 2MASS ~\cite{a2}, SDSS ~\cite{a3}. Space based observations are taken from MSX ~\cite{a4}, IRAS ~\cite{a5}. The LAMOST, WISE, MSX, and 2MASS catalogues are chosen as the data source. LAMOST is a spectroscopic survey with the wavelength coverage of 0.37 - 0.9 $\upmu$m, which located at the Xinglong Observatory in Hebei, China ~\cite{z}. SDSS provides photometric as well as spectroscopic data ~\cite{a6}. The WISE catalogue provides detection data in four infrared bands, corresponding to the central wavelengths of 3.4, 4.6, 12, and 22 $\upmu$m ~\cite{a1}. 2MASS provides the flux density of three bands, corresponding to the central wavelengths of 1.235, 1.662, and 2.159 $\upmu$m ~\cite{a2}.

The three catalogues all give information of the stellar magnitude, and the energy information should be converted into flux density, which is required by the model. The relationship between the brightness of two celestial bodies and magnitude is given by Pogson's equation:

\begin{equation}
\label{eq:2.10}
{m_2} - {m_1} =  - 2.5\log \frac{{{E_2}}}{{{E_1}}}                   
\end{equation}

where $\emph{m}_1$ and $\emph{m}_2$ are the stellar magnitudes, and $\emph{E}_1$ and $\emph{E}_2$ are the flux densities. The photometric zero point is usually adopted in the application of this formula. The flux density $\emph{E}_1$ is the photometric zero point corresponding to the condition $\emph{m}_1$ = 0. The stellar flux density $\emph{E}_2$ can be calculated according to the given stellar magnitude $\emph{m}_2$.

\begin{table}[!htbp]
\centering
\begin{tabular}{cccccc}
\hline
\multirow{6}*{LAMOST} & FilterID & $\lambda_{eff}$ ($\upmu$m) & $\lambda_{min}$ ($\upmu$m) & $\lambda_{max}$ ($\upmu$m) & ${f_\lambda}$ ($\times$10$^{-8}$ Wm$^{-2}$ $\upmu$m$^{-1}$)\\
\cline{2-6}
               & $\emph{u}$ & 0.356	& 0.32	&0.38	&3.67\\
               & $\emph{g}$ & 0.483	& 0.41	&0.55	&5.11\\
               & $\emph{r}$ & 0.626	& 0.555	&0.695	&2.40\\
               & $\emph{i}$ & 0.767	& 0.695	&0.845	&1.82\\
               & $\emph{z}$ & 0.910	& 0.85	&0.97	&0.783\\
\hline
\multirow{5}*{WISE} & FilterID  & $\lambda_{eff}$ ($\upmu$m) & $\lambda_{min}$ ($\upmu$m) & $\lambda_{max}$ ($\upmu$m) & ${f_\nu}$ ($\times$10$^{-26}$ Wm$^{-2}$ Hz$^{-1}$)\\
\cline{2-6}
               & $\emph{W}1$&3.3526 &2.7541	&3.8724	&309.5\\
               & $\emph{W}2$&4.6028 &3.9633	&5.3414	&171.8\\
               & $\emph{W}3$&11.5608&7.4430	&17.2613&31.7\\
               & $\emph{W}4$&22.0883&19.5201	&27.9107&8.4\\
\hline
\multirow{4}*{2MASS} & Filter ID&$\lambda_{eff}$ ($\upmu$m)&$\lambda_{min}$ ($\upmu$m)&$\lambda_{max}$ ($\upmu$m)&${f_\nu}$ ($\times$10$^{-26}$ Wm$^{-2}$ Hz$^{-1}$)\\
\cline{2-6}
               & $\emph{J}$ &1.2350&	1.153795&	1.316205&	1594.0\\
               & $\emph{H}$ &1.6620&	1.53653 &	1.78747 &	1024.0\\
               & $\emph{K}_s$&2.1590&	2.028055&	2.289945&	666.8\\
\hline
\end{tabular}
\caption{\label{tab:1} Photometric zone points of different catalogues (LAMOST, WISE, 2MASS).}
\end{table}

The photometric zero point is different for different wavelength bands. The specific band ranges and photometric zone points of LAMOST are similar to SDSS catalogue. The specific band ranges and photometric zone points of the LAMOST ~\cite{a7}, WISE ~\cite{a8}, and 2MASS ~\cite{a9} catalogues are presented in table~\ref{tab:1}.

The unit in the website is Jy, the convention relationship is 1 Jy=10$^{-26}$ Wm$^{-2}$ Hz$^{-1}$. The unit of $\emph{f}_\lambda$ can be obtained from the reference ~\cite{a7} and the following conversion relationship: 1*10$^{-9}$cgs/ \AA = 1*10$^{-8}$Wm$^{-2}$ $\upmu$m$^{-1}$. The last column can be obtained for LAMOST catalogues. For WISE and 2MASS catalogues, the last column can be obtained from the following relationship: 1Jy=10$^{-26}$Wm$^{-2}$ Hz$^{-1}$.

The MSX catalogue contains 177 860 stars and lists the average flux density estimates of these stars in six bands, which are 6.8 - 10.8, 4.22 - 4.36, 4.24 - 4.45, 11.1 - 13.2, 13.5 - 15.9, and 18.2 - 25.1 $\upmu$m ~\cite{y}. The data "version 2.3 of the MSX point source catalog" was used in this paper.

The flux density given in the MSX catalogue should be converted to radiative intensity. The following formula describes the relationship of the flux density between wavelength and frequency.

\begin{equation}
\label{eq:2.11}
{E_{{\lambda _1} - {\lambda _2}}}{\rm{ = }}\int_{{\lambda _1}}^{{\lambda _2}} {{E_\lambda }{\rm{d}}\lambda  = } \int_{c/{\lambda _1}}^{c/{\lambda _2}} {{E_\gamma }{\rm{d}}\gamma } 
\end{equation}

\subsection{SPSO algorithm}
\label{sec:spso}
The effective temperatures and angular parameters are calculated by the inversion method according to the flux density of the given catalogues. Because of the non-linearity of the equation, a numerical method is used to solve the problem instead of an analytic solution. More specifically, the SPSO algorithm is used. The specific idea behind SPSO is that, for the question of being retrieved, each possible solution is expressed as a particle in the population, and each particle has its own position and velocity. All particles move at a certain velocity in the solution space and find the global optimal value by following a fitness optimal value determined by the objective function ~\cite{a10, a11, a12, a13}.

The optimization function is

\begin{equation}
\label{eq:2.12}
Fitnes{s_i}{\rm{ = }}\sqrt {{{\left( {\left( {{E_{ipa}} - {E_{ipb}}} \right)/{E_{ipa}}} \right)}^2}}                      
\end{equation}

where $\emph{E}_{ipa}$ represents the initial value of the inversion, and $\emph{E}_{ipb}$ represents the value for particle $\emph{i}$.

Effective temperatures are assumed within 1000 - 20000 K and detected angular parameters are between 1.0$\times$10$^{-21}$ and 1.0$\times$10$^{-16}$. This determined the boundary of the solution space.

Mathematical description of PSO is as follows. The number of particles is $\emph{M}$ in a $\emph{D}$-dimensional search space, and the spatial position of each particle represents a potential solution. Position vector for particle $\emph{i}$ is  ${X_i} = \left( {{x_{i1}},{x_{i2}}, \cdot  \cdot  \cdot {x_{iD}}} \right)$, and velocity vector is ${V_i} = \left( {{v_{i1}},{v_{i2}}, \cdot  \cdot  \cdot {v_{iD}}} \right)$ . The best position that this particle has experienced is ${P_i} = \left( {{p_{i1}},{p_{i2}}, \cdot  \cdot  \cdot {p_{iD}}} \right)$, and is denoted by $\emph{P}_{best}$. The corresponding best position of all particles is ${P_g} = \left( {{p_{g1}},{p_{g2}}, \cdot  \cdot  \cdot {p_{gD}}} \right)$ and is denoted by $\emph{g}_{best}$. The particle velocity depends on the personal best and global best, and it is given by

\begin{equation}
\label{eq:2.13}
{V_i}\left( {t + 1} \right) = w{V_i}\left( t \right) + {c_1}{r_1}\left[ {{P_i}\left( t \right) - {X_i}\left( t \right)} \right] + {c_2}{r_2}[{P_g}\left( t \right) - {X_i}\left( t \right)]  
\end{equation}

Here, $\emph{t}$ is the current iteration, $\emph{w}$ is the inertia weight, $\emph{c}_1$ and $\emph{c}_2$ are constant accelerations, and $\emph{r}_1$ and $\emph{r}_2$ are random numbers in [0, 1]. The new location of $\emph{X}_i$ is

\begin{equation}
\label{eq:2.14}
{X_i}\left( {t + 1} \right) = {X_i}\left( t \right){\rm{ + }}{V_i}\left( {t{\rm{ + }}1} \right)                     
\end{equation}

This formula reduces the global search capability, and increases the local search capability. So, if $\emph{X}_j(t)$ = $\emph{P}_j$ =$\emph{P}_g$, particle $\emph{j}$ will be "flying" at the velocity zero. To improve the global search capability, we conserve the current best position of the swarm $\emph{P}_g$ and the $\emph{j}$'s best position $\emph{P}_j$, then giving a new particle $\emph{j}$'s position $\emph{X}_j(t+1)$, and other particles are manipulated according to (15), thus the global search capability is enhanced.

\begin{figure}[!htbp]
\setlength{\abovecaptionskip}{0.cm}
\setlength{\belowcaptionskip}{-0.cm}
    \centering
    \begin{tabular}{cc} 
        \includegraphics[width=7.0cm, height=6.0cm, angle=0]{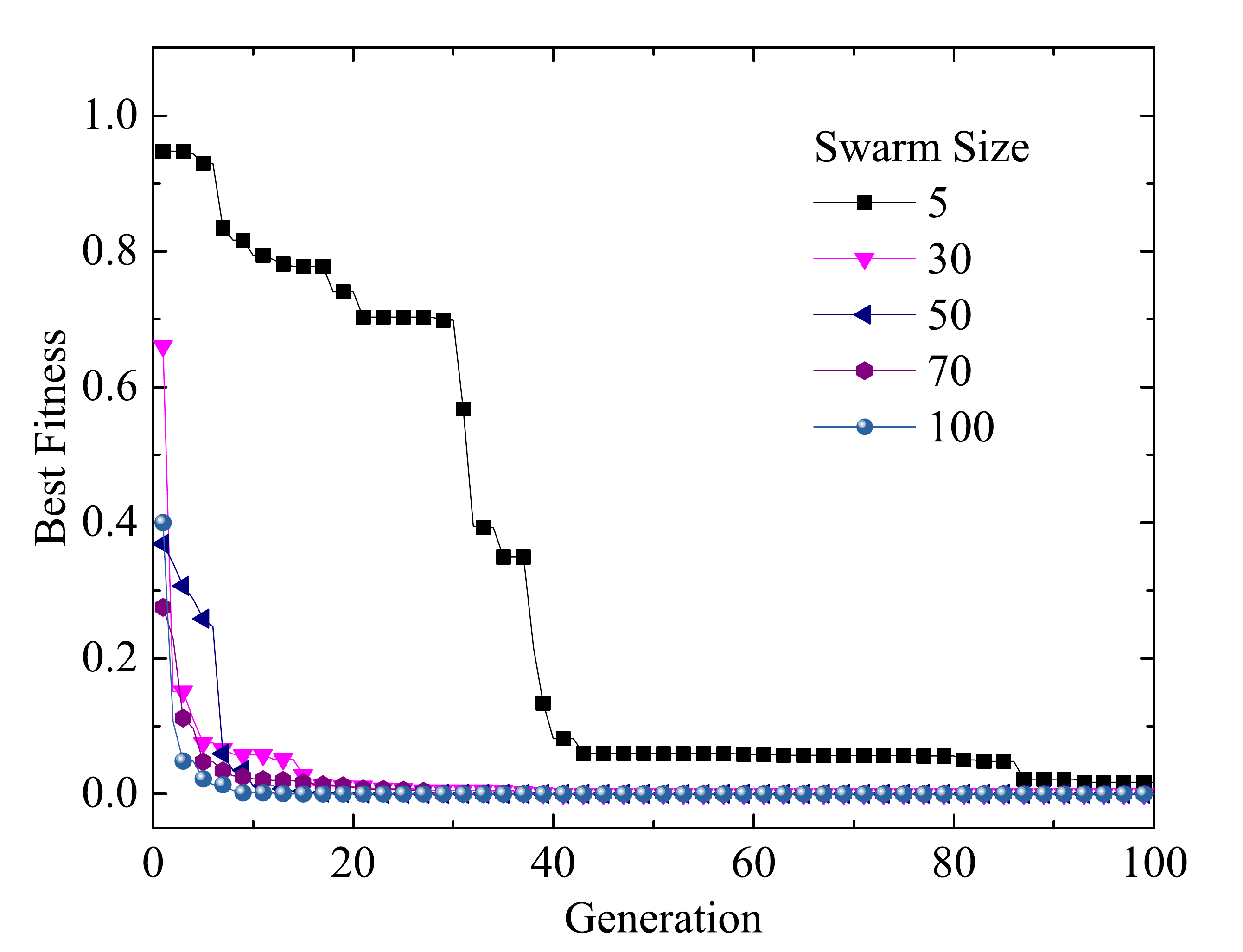} & \includegraphics[width=7.0cm, height=6.0cm, angle=0]{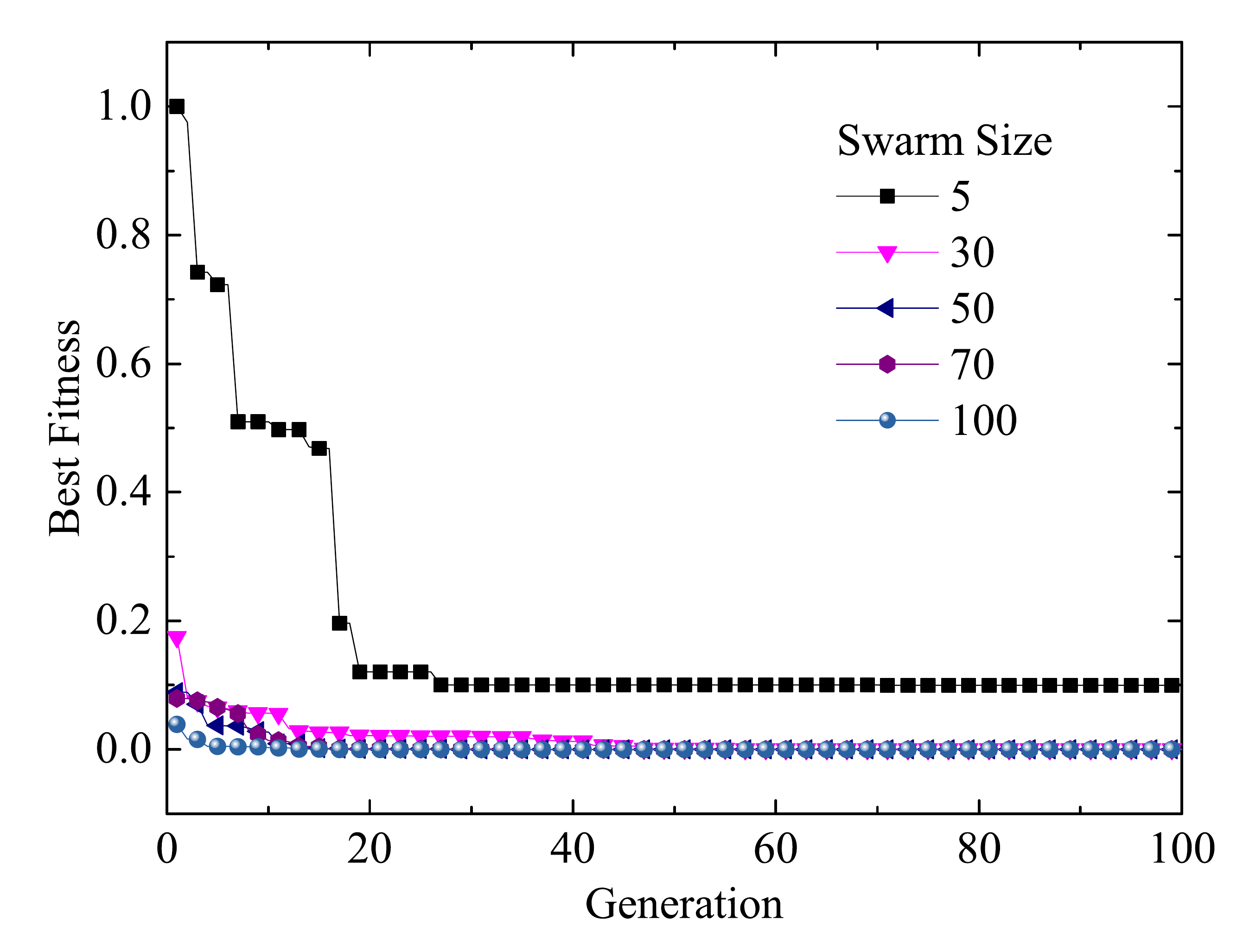}
        \\
        (a) LAMOST & (b) WISE
        \\
        \includegraphics[width=7.0cm, height=6.0cm, angle=0]{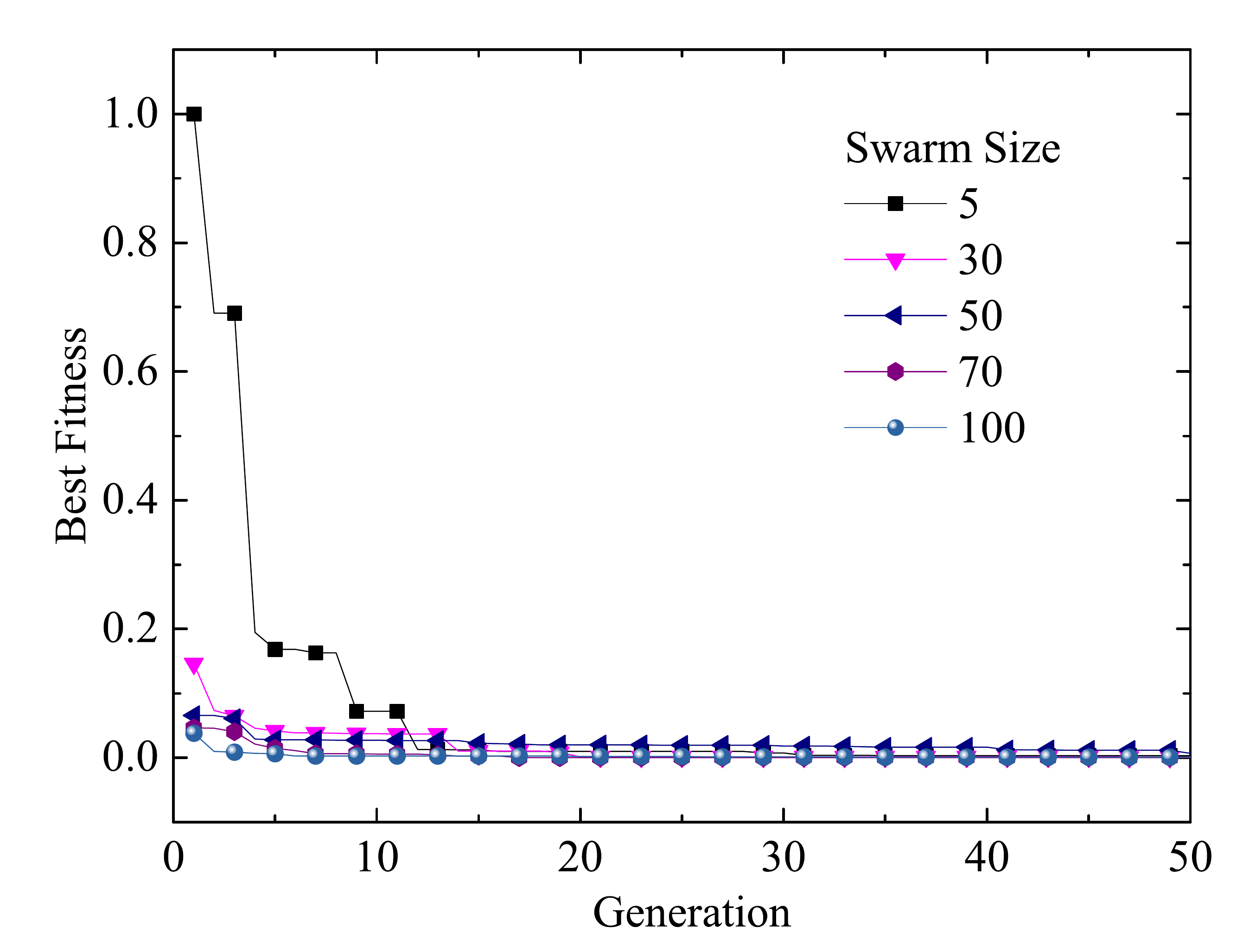} & \includegraphics[width=7.0cm, height=6.0cm, angle=0]{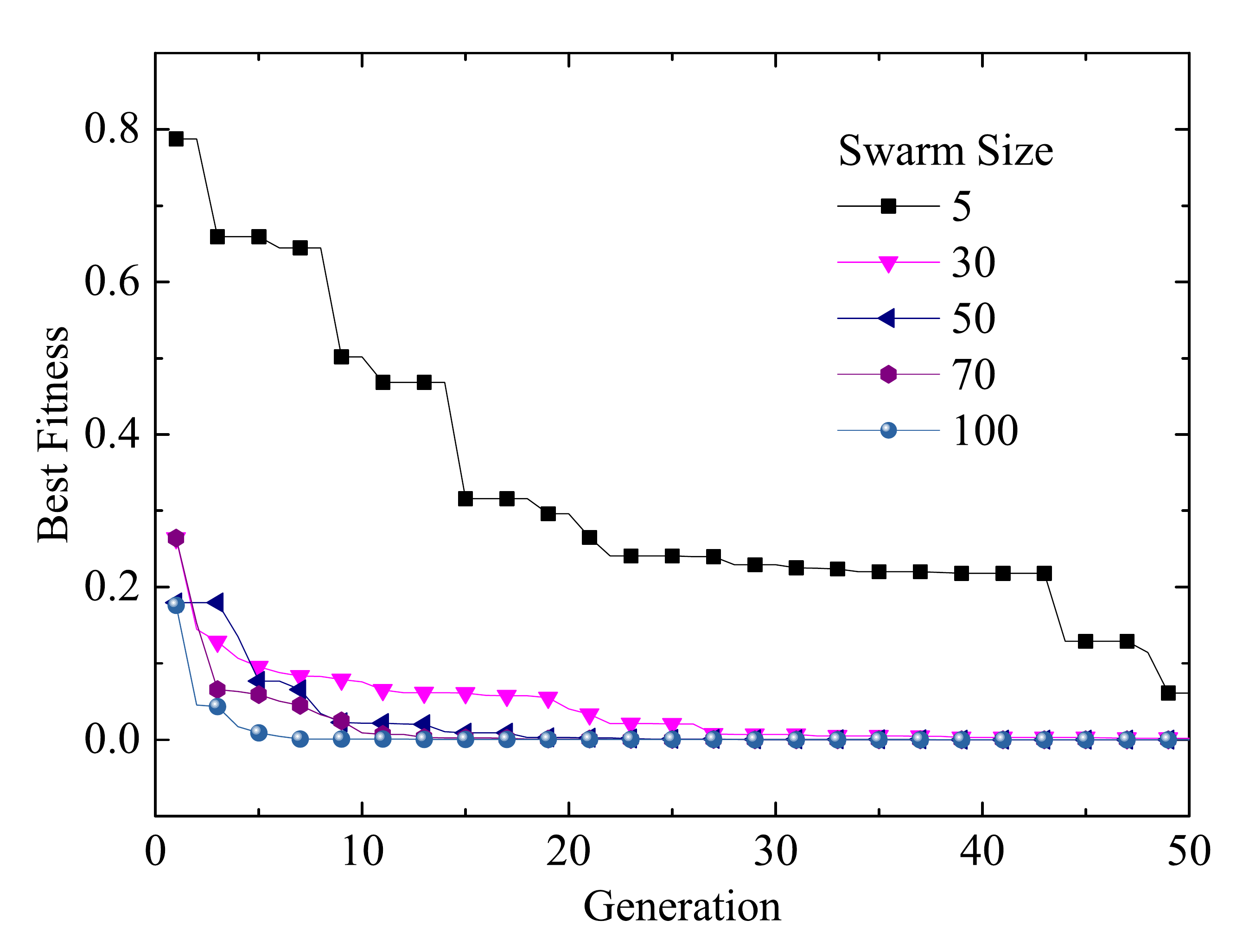}
        \\
        (c) MSX & (d) 2MASS
    \end{tabular}
    \caption{Fitness values as a function of generation with different particle numbers for (a) LAMOST, (b) WISE, (c) MSX, and (d) 2MASS.}
    \label{fig:2}
\end{figure}

The flux density is obtained according to Planck's law in the case that the stellar temperature is 5000 K. The temperature is assumed to verify the algorithm. The system fitness values are calculated based on the given flux density. As the detection bands of the LAMOST, WISE, MSX, and 2MASS catalogues are different, the SPSO algorithm needs to be verified. The fitness values of the four star catalogues as a function of generation calculated with different particle numbers are presented in figure~\ref{fig:2}.

According to the figure, the fitness values gradually decrease with the calculation number with different particle numbers for each detection band. When the particle number is 5, the fitness value decreases slowly, but when the particle number reaches 100, the fitness value decreases rapidly. Considering the calculation time and efficiency, we choose the number of particles as 50.

\section{Results and discussion}
\label{sec:results}
The SPSO algorithm is used to calculate $T_{eff}$ and angular parameters using given flux densities of star catalogues. Compared with the star catalogue data, the best band combination of the inversion calculation is determined. The relative error of flux density and the variance of the relative error of each band are calculated using the optimal band combination and flux density of the given bands. The applicable range of the model is obtained.

\subsection{Optimal band combinations}
Errors exist in the measured flux density or luminosity because of the influence of the structure of the ground or space telescope detectors and their measurements ~\cite{f}. The errors cause the inversion result of flux density to be inaccurate. To weigh the accuracy of the results, the mean standard deviation (mean normalized bias) is used:

\begin{equation}
\label{eq:3.1}
M = \frac{1}{3}\sum\limits_{i = 1}^3 {\left( {{{\left| {{X_i} - {D_i}} \right|} \mathord{\left/
 {\vphantom {{\left| {{X_i} - {D_i}} \right|} {{D_i}}}} \right.
 \kern-\nulldelimiterspace} {{D_i}}}} \right)}                      
\end{equation}

where $\emph{M}$ is the mean standard deviation, $\emph{X}_i$ is the inversion result, $\emph{D}_i$ is the detected value, and $\emph{n}$ is the total number of calculated results.

\begin{table}[!htbp]
\centering
\begin{tabular}{ccc}
\hline
\multirow{7}*{LAMOST} &  Scheme	&Band combinations ($\upmu$m)\\
\cline{2-3}
               &   A1	&0.32 - 0.38, 0.41 - 0.55, 0.555 - 0.695\\
               &   B1	&0.32 - 0.38, 0.41 - 0.55, 0.695 - 0.845\\
               &   C1	&0.32 - 0.38, 0.555 - 0.695, 0.85 - 0.97\\
               &   D1	&0.41 - 0.55, 0.555 - 0.695, 0.695 - 0.845\\
               &   E1	&0.41 - 0.55, 0.695 - 0.845, 0.85 - 0.97\\
               &   F1	&0.555 - 0.695, 0.695 - 0.845, 0.85 - 0.97\\
\hline
\multirow{5}*{WISE} &  Scheme	&Band combinations ($\upmu$m)\\
\cline{2-3}
               &   A2	&2.754 - 3.872, 3.963 - 5.341, 7.443 - 17.261\\
               &   B2	&2.754 - 3.872, 3.963 - 5.341, 19.520 - 27.911\\
               &   C2	&2.754 - 3.872, 7.443 - 17.261, 19.520 - 27.911\\
               &   D2	&3.963 - 5.341, 7.443 - 17.261, 19.520 - 27.911\\
\hline
\multirow{7}*{MSX} &  Scheme	&Band combinations ($\upmu$m)\\
\cline{2-3}
               &   A3	&6.8 - 10.8, 4.22 - 4.36, 4.24 - 4.45\\
               &   B3	&6.8 - 10.8, 4.22 - 4.36, 13.5 - 15.9\\
               &   C3	&6.8 - 10.8, 4.22 - 4.36, 18.2 - 25.1\\
               &   D3	&6.8 - 10.8, 13.5 - 15.9, 18.2 - 25.1\\
               &   E3	&4.22 - 4.36, 4.24 - 4.45, 11.1 - 13.2\\
               &   F3	&11.1 - 13.2, 13.5 - 15.9, 18.2 - 25.1\\
\hline
\end{tabular}
\caption{\label{tab:2} Band combinations.}
\end{table}

Two unknowns ($T_{eff}$ and angular parameters) need to be solved in the problem. If only two bands are selected, the formula will be ill-posed. It would lead to multiple solutions. If the bands are more than three, the formula is overdetermined, which leads to no solution. The flux densities of the three bands should be selected as the input source for the inversion requirement of the SPSO algorithm. Wavelength ranges and intervals are different for all bands. Proper band combinations are selected to avoid small ranges and close interval. The accuracy of the inversion results is different for each band using different band combinations as the input source. This effect is analysed and the best band combination is selected.

The flux density is calculated with different band combinations for the different star catalogues as follows. LAMOST has six band combinations, WISE has four band combinations, and MSX has six band combinations, as shown in table~\ref{tab:2}. 2MASS only has one combination for three bands.

\begin{figure}[!htbp]
\setlength{\abovecaptionskip}{0.cm}
\setlength{\belowcaptionskip}{-0.cm}
    \centering
    \begin{tabular}{cc} 
        \includegraphics[width=7.0cm, height=6.0cm, angle=0]{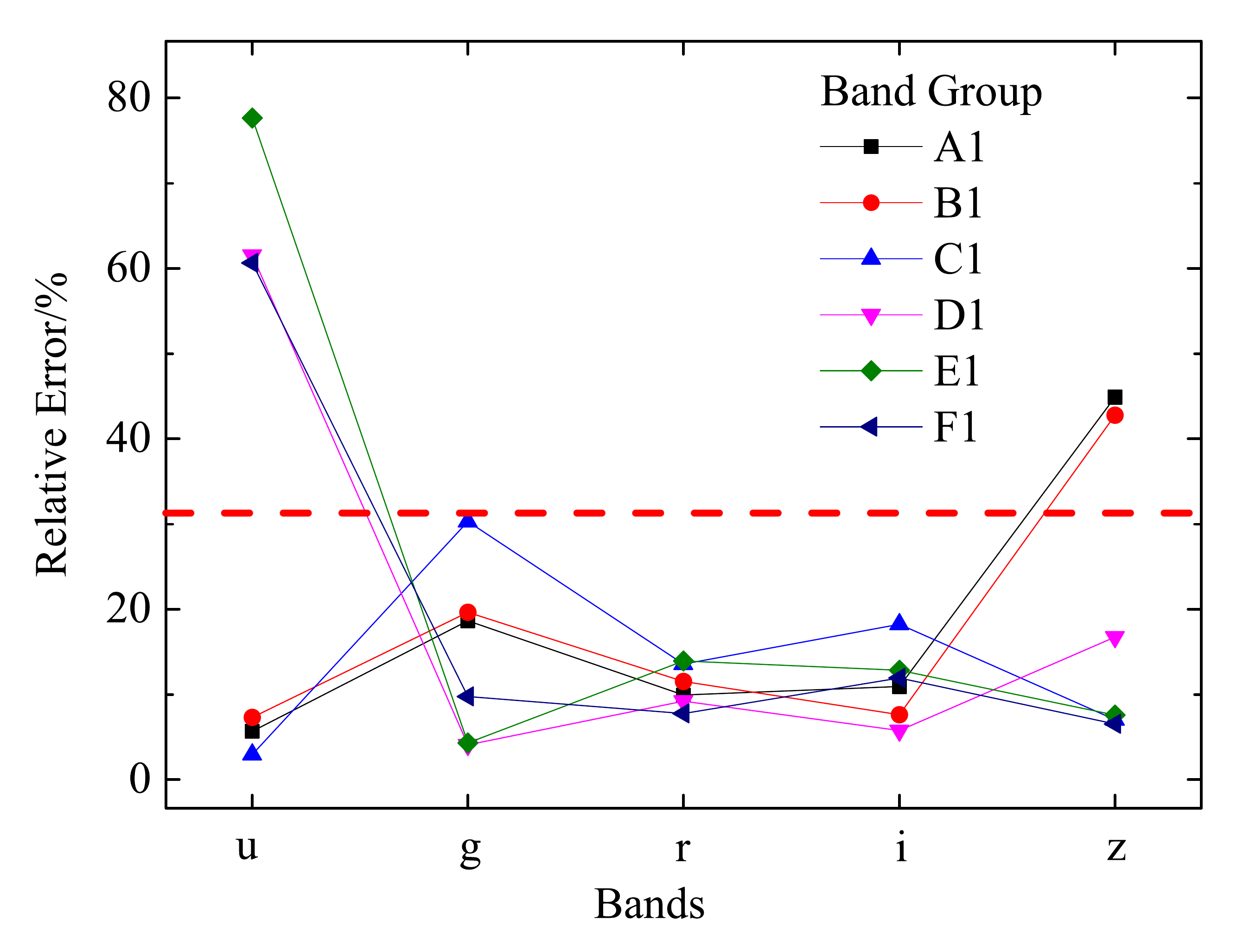} & \includegraphics[width=7.0cm, height=6.0cm, angle=0]{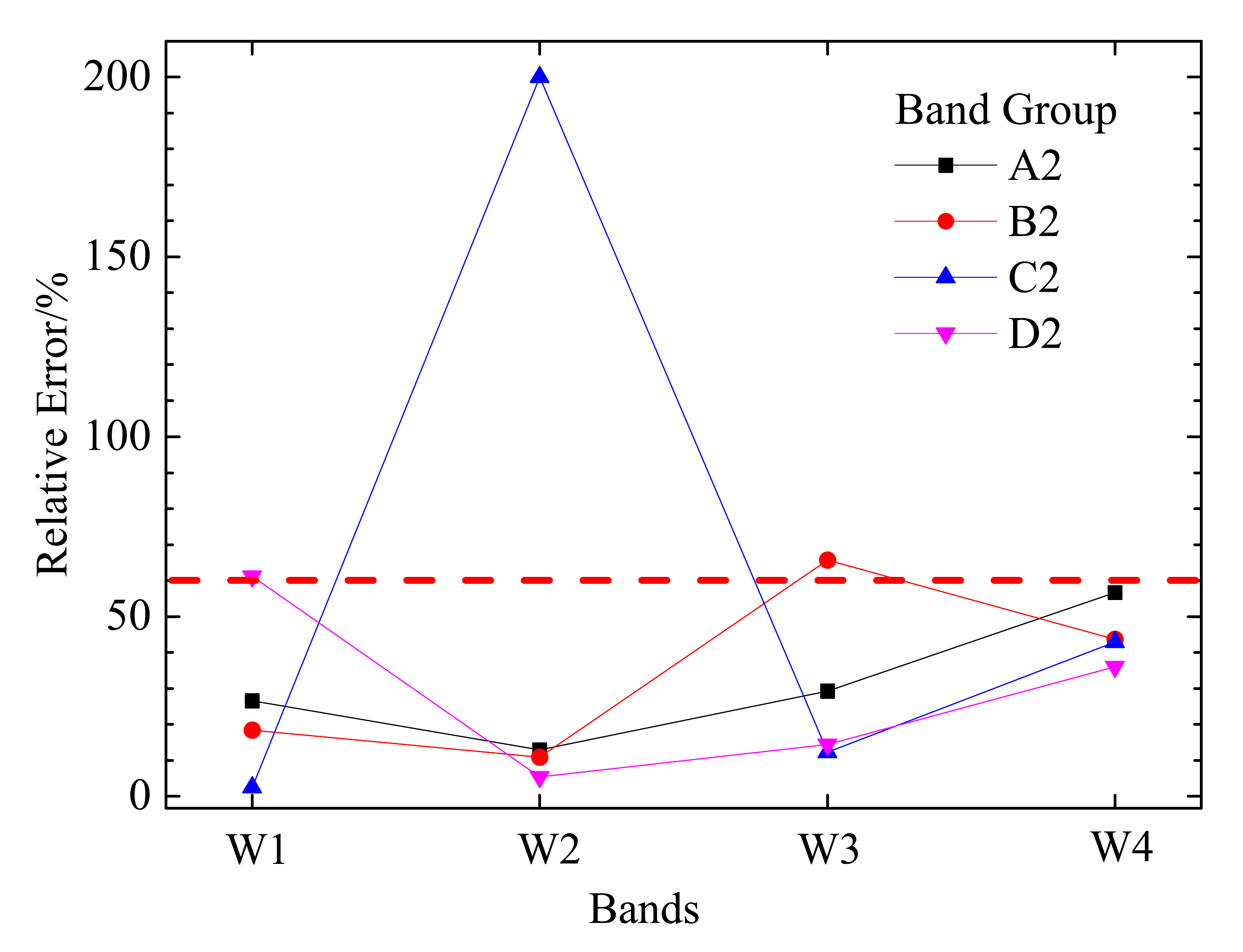}
        \\
        (a) LAMOST & (b) WISE
        \\
        \includegraphics[width=7.0cm, height=6.0cm, angle=0]{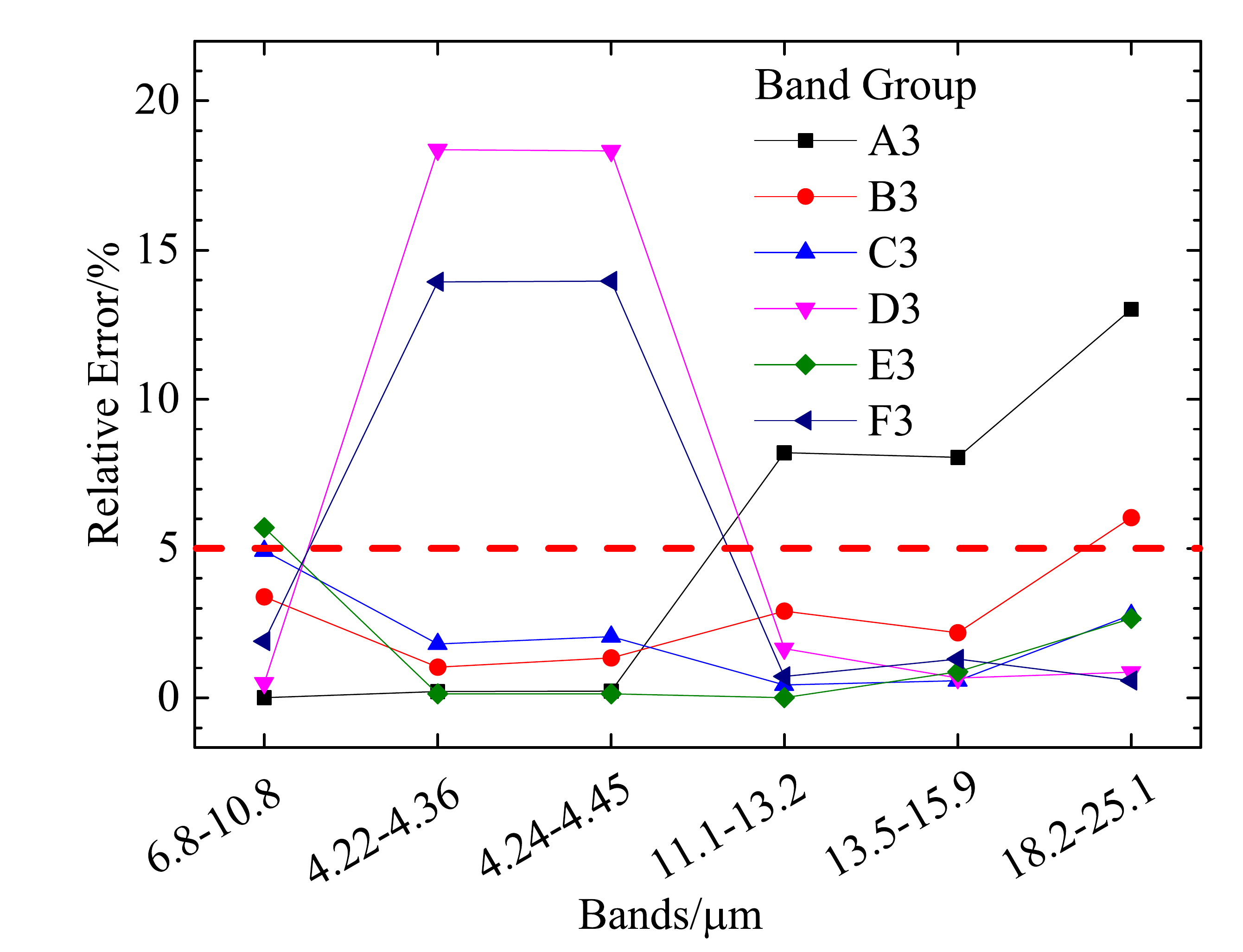} & \includegraphics[width=7.0cm, height=6.0cm, angle=0]{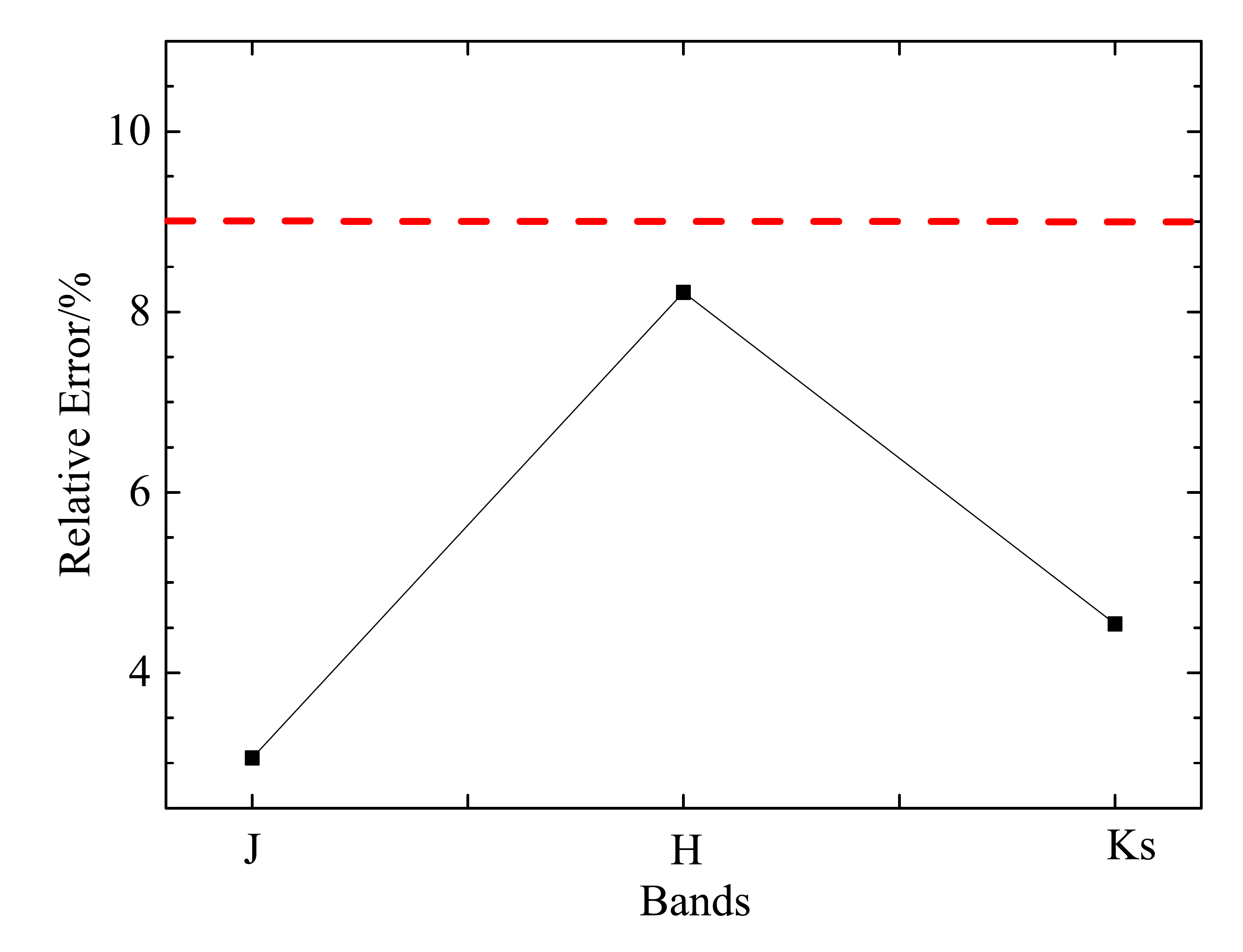}
        \\
        (c) MSX & (d) 2MASS
    \end{tabular}
    \caption{Relative errors as a function of wavelength in different band combinations for (a) LAMOST, (b) WISE, (c) MSX, and (d) 2MASS. The red dotted line is a criterion of minimum relative error to select the best combination.}
    \label{fig:3}
\end{figure}

Figure~\ref{fig:3} presents the results of the inversion calculation according to the flux density of the star catalogues with different band combinations.

The criterion of the relative error for the LAMOST catalogue, which is less than 31 per cent for all bands as indicated by the dashed line in figure~\ref{fig:3}(a), is selected. The errors in the $\emph{u}$ band are more than 60 per cent when the combinations of D1, E1, and F1 are selected. The errors in the $\emph{z}$ band are more than 40 per cent when the combinations of A1 and B1 are selected. These are not suitable as the inversion input band combinations. For the C1 combination, the relative error in the five bands is low and satisfies the requirement of less than 31 per cent.

Similarly, the relative error is less than 60 per cent for all bands of the WISE catalogues as the criterion, as shown in figure~\ref{fig:3}(b). For the combinations of B2, C2, and D3, the errors are high and not suitable as the inversion input band combinations. The relative errors of all bands are less than 60 per cent for the A2 combination, satisfying the requirement.

For the MSX catalogues, as shown in figure~\ref{fig:3}(c), a relative error of less than 5 per cent for all bands is considered as the criterion. When the combinations of D3 and F3 are selected, the errors of the 4.22 - 4.36 and 4.24 - 4.45 $\upmu$m bands are more than 13 per cent. The errors in the 18.2 - 25.1 $\upmu$m band are more than 5 per cent for the combinations of A3 and B3. The error in the 6.8 - 10.8 $\upmu$m band is more than 5 per cent for the combination of E3. These are not suitable as the input band combinations. For the C3 combination, the relative errors in the six bands are all less than 5 per cent, satisfying the requirement.

For the 2MASS catalogues, as shown in figure~\ref{fig:3}(d), given the three bands, the relative error is less than 9 per cent, so it can be used as the input source of the inversion calculation.

Therefore, the best band combinations are C1, A2, and C3 for the LAMOST, WISE, and MSX catalogues, respectively. The best band combinations of the four star catalogues obtained above are used to calculate the flux density in the following.

\subsection{Stellar flux density calculation}
According to the radiative energy information given by the star catalogues, the flux densities of the other corresponding bands are calculated by the best band combination obtained in Sec. 3.1. The flux densities are obtained using eq. \eqref{eq:2.9}. The variance of the relative error is calculated. The average standard deviation is calculated by eq. \eqref{eq:2.13}, and relative errors of the flux density of the four star catalogues are obtained. The variance S2 of the relative error is calculated using the following modified sample variance formula:

\begin{equation}
\label{eq:3.2}
{S^2} = \frac{1}{{n - 1}}\sum\limits_{i = 1}^n {{{\left( {{{\left| {{X_i} - {D_i}} \right|} \mathord{\left/
 {\vphantom {{\left| {{X_i} - {D_i}} \right|} {{D_i}}}} \right.
 \kern-\nulldelimiterspace} {{D_i}}} - M} \right)}^2}}                                          
\end{equation}

The flux densities of 964 stars are calculated using the band combination C1 in the LAMOST catalogue. The relative error of each band is summarized in table~\ref{tab:3}. The relative errors of the $\emph{u}$, $\emph{r}$, and $\emph{z}$ bands are all less than 15 per cent. Especially for the $\emph{u}$ band, the relative error reaches 2.956 per cent, which satisfies the requirement better. The relative error is large for the $\emph{g}$ band, reaching 30.325 per cent.
\begin{table}[!htbp]
\centering
\begin{tabular}{cccccc}
\hline
Band ($\upmu$m)	&$\emph{u}$&	$\emph{g}$&	$\emph{r}$&	$\emph{i}$&	$\emph{z}$\\
\hline
Relative error (\%)	&2.956	&30.325	&13.566	&18.233	&7.057\\
Variance ($\times$10$^{-3}$)	&1.348	&375	&86.99	&140	&1.948\\
\hline
\end{tabular}
\caption{\label{tab:3} Relative errors and variances of LAMOST catalogue.}
\end{table}

\begin{figure}[!htbp]
\setlength{\abovecaptionskip}{0.cm}
\setlength{\belowcaptionskip}{-0.cm}
   \centering
   \includegraphics[width=14.0cm, height=12.0cm, angle=0]{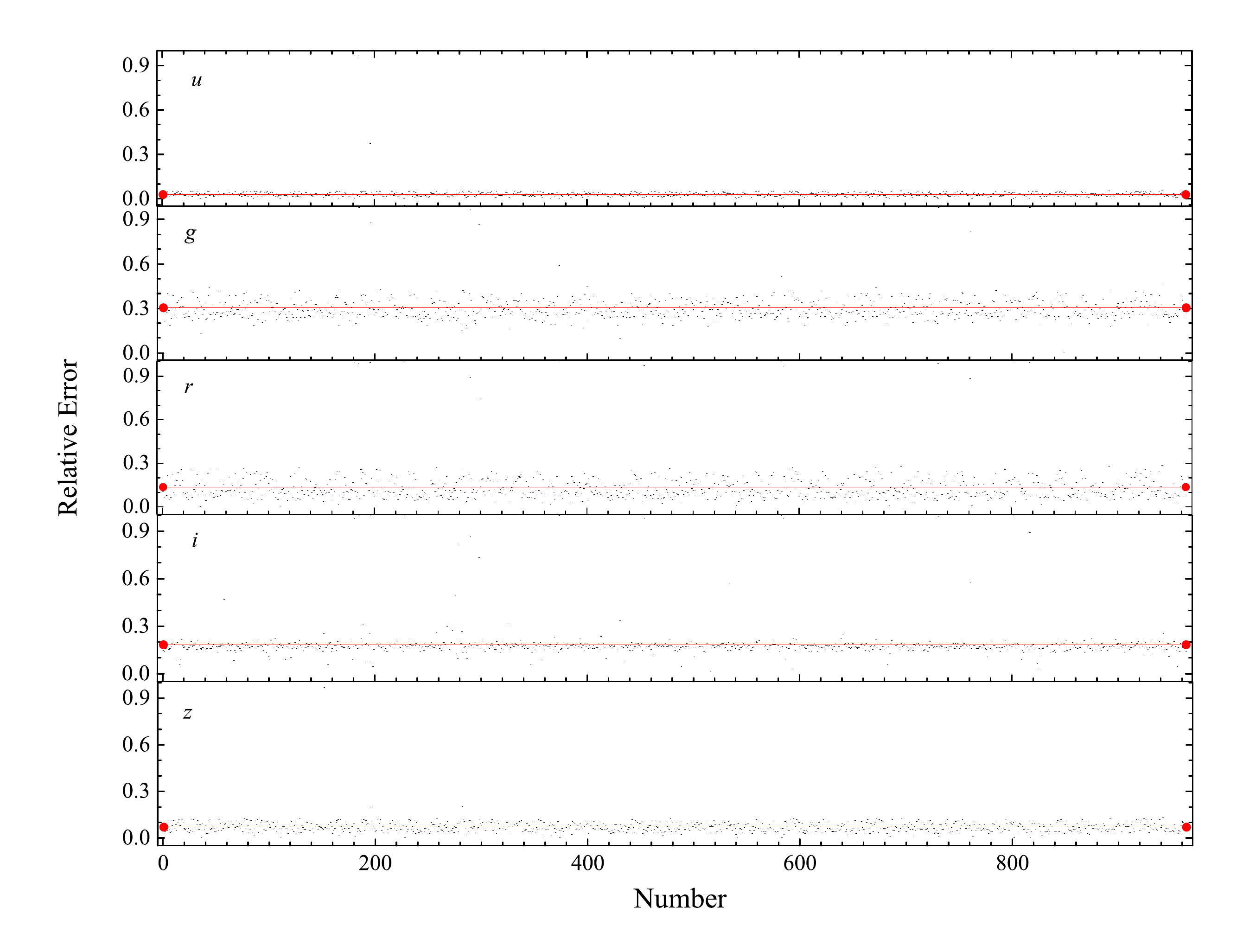}
   \caption{Relative error variation (LAMOST).}
   \label{fig:4}
   \end{figure}

For the five bands of LAMOST, the specific variation of relative error is presented in figure~\ref{fig:4}. The errors are relatively concentrated and the variance is small. The variance reaches 1.348$\times$10$^{-3}$, 86.99$\times$10$^{-3}$, and 1.948$\times$10$^{-3}$ for the $\emph{u}$, $\emph{r}$, and $\emph{z}$ bands, respectively. The relative errors are scattered and the variances are larger for the $\emph{g}$ and $\emph{i}$ bands. However, none of the bands are larger than 0.5, which indicates that in overall the errors are small and concentrated. They can be used to invert the flux density using the corresponding catalogue.

The flux densities of 489 stars in the WISE catalogue are calculated using band combination A2. The errors of the four bands are summarized in table~\ref{tab:4}. The errors of the $\emph{W}1$, $\emph{W}2$, and $\emph{W}3$ bands are all less than 30 per cent, especially for the $\emph{W}2$ band, and the relative error is 12.897 per cent, which satisfies the requirement better. However, for the $\emph{W}4$ band, the error is large, reaching 56.427 per cent.
\begin{table}[!htbp]
\centering
\begin{tabular}{ccccc}
\hline
Band ($\upmu$m)&	$\emph{W}1$&	$\emph{W}2$ &	$\emph{W}3$&	$\emph{W}4$\\
\hline
Relative error (\%)	&26.354	&12.897&	29.325&	56.427\\
Variance	&0.378	&9.728 &0.491&	1.398\\
\hline
\end{tabular}
\caption{\label{tab:4} Relative errors and variances of WISE catalogue.}
\end{table}

\begin{figure}[!htbp]
\setlength{\abovecaptionskip}{0.cm}
\setlength{\belowcaptionskip}{-0.cm}
\centering 
\includegraphics[width=14.0cm, height=12.0cm, angle=0]{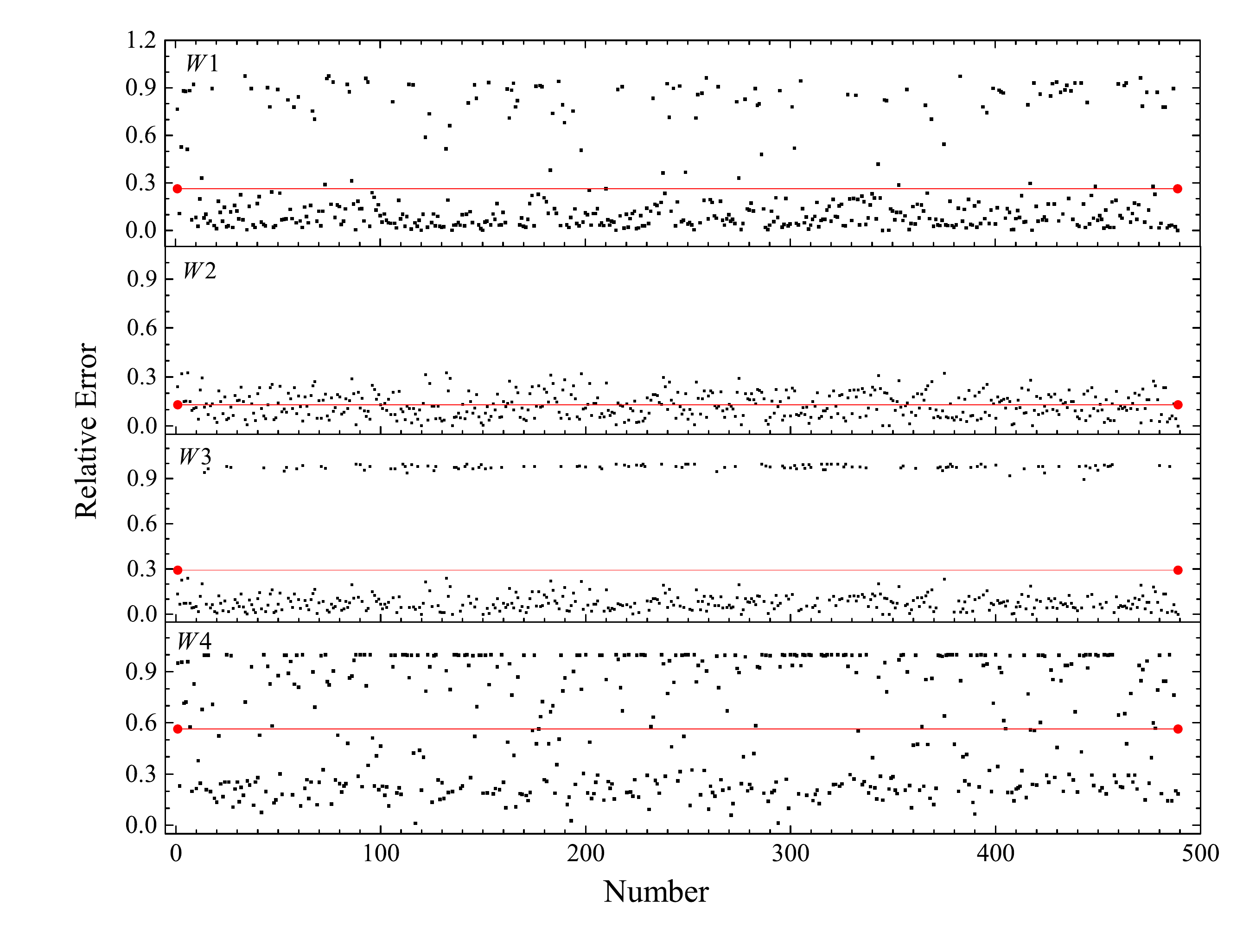}
\caption{Relative error variation (WISE).}
\label{fig:5}
\end{figure}

The specific variations in the relative errors are presented in figure~\ref{fig:5} for the four bands. The errors are relatively concentrated, and the variances are small for the $\emph{W}2$ band, reaching 9.728. The errors are scattered and the variances are larger for the other three bands. It is clear that the flux densities need to be revised for the WISE catalogue to obtain accurate results.

The flux densities of 1000 stars are calculated by the band combination C3 in the MSX catalogue. The error of each band is summarized in table~\ref{tab:5}, which indicate that they are all less than 5 per cent, satisfying the requirement better. Especially, the relative errors of the 11.1 - 13.2 and 13.5 - 15.9 $\upmu$m bands are 0.432 and 0.569 per cent, respectively.
\begin{table}[!htbp]
\centering
\begin{tabular}{ccccccc}
\hline
Band ($\upmu$m)& 6.8 - 10.8 &4.22 - 4.36& 4.24 - 4.45&11.1 - 13.2&13.5 - 15.9&18.2 - 25.1\\
\hline
Relative error (\%)&4.924&	1.798	&2.045	&0.435	&0.567	&2.774\\
Variance ($\times$10$^{-7}$)	&97090	&4.863	&7.777	&3.658	&1287	&3.837\\
\hline
\end{tabular}
\caption{\label{tab:5} Relative errors and variances of MSX catalogue.}
\end{table}

\begin{figure}[!htbp]
\setlength{\parskip}{0\normalbaselineskip} 
\setlength{\textfloatsep}{0.3\normalbaselineskip}
\setlength{\abovecaptionskip}{0.cm}
\setlength{\belowcaptionskip}{-0.cm}
\centering 
\includegraphics[width=14.0cm, height=12.0cm, angle=0]{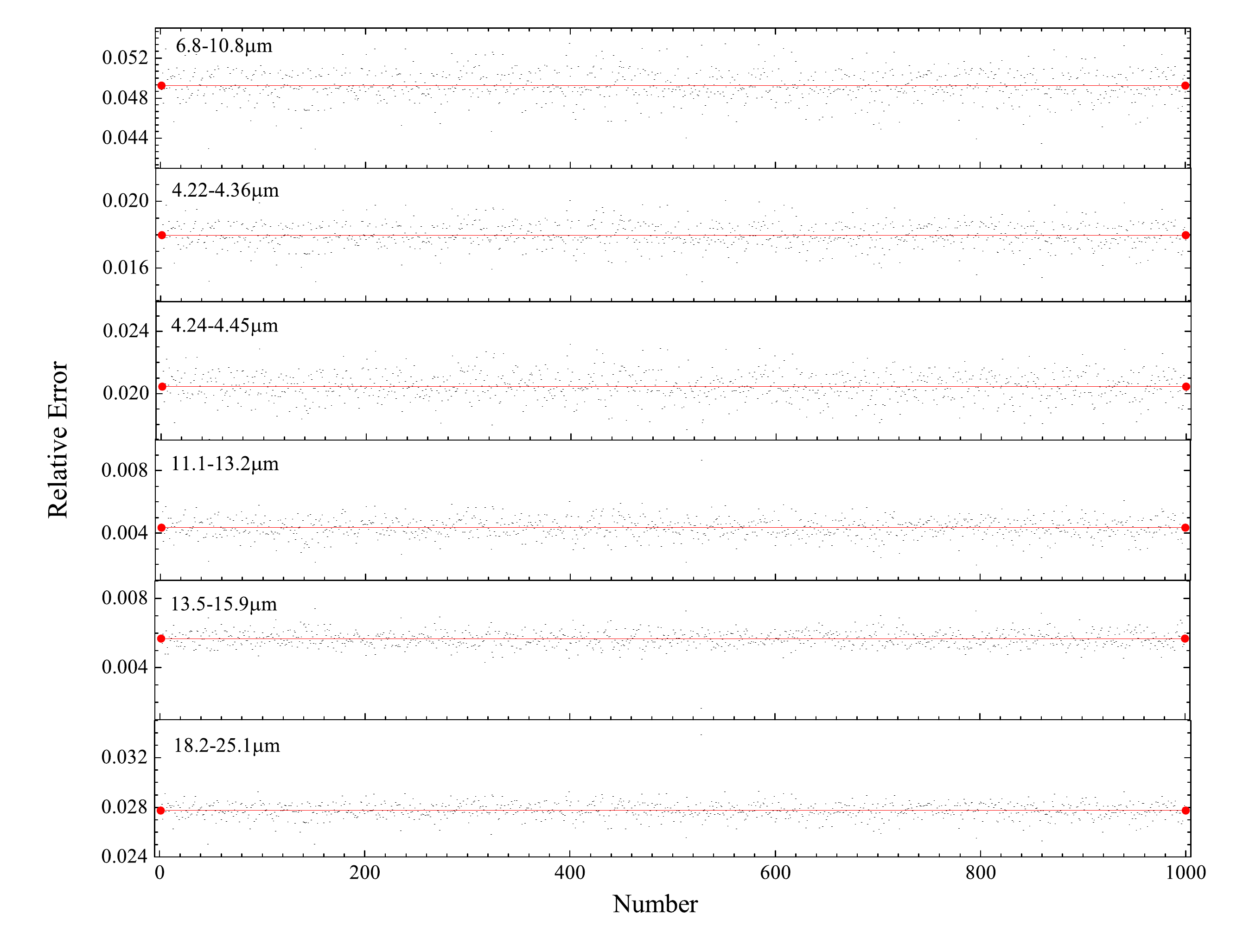}
\caption{ Relative error variation (MSX).}
\label{fig:6}
\end{figure}

The specific variations in the relative errors of the six bands are presented in figure~\ref{fig:6}. The errors of all bands are relatively concentrated and the variances are small, which can be used to invert the flux densities of the corresponding bands.

The flux density of 403 stars in the 2MASS catalogue is calculated. The error of each band is summarized in table~\ref{tab:6}, which indicates that the they are all less than 10 per cent, satisfying the requirement better. Especially, for the J band, the relative error reaches 3.058 per cent.
\begin{table}[!htbp]
\centering
\begin{tabular}{cccc}
\hline
Band ($\upmu$m)&$\emph{J}$&$\emph{H}$&$\emph{K}_s$\\
\hline
Relative error (\%) &	3.058 &	8.216 &	4.541\\
Variance ($\times$10$^{-3}$) &	1.408 &	25.43 &	3.667\\
\hline
\end{tabular}
\caption{\label{tab:6} Relative errors and variances of 2MASS catalogue.}
\end{table}

\begin{figure}[!htbp]
\setlength{\parskip}{0\normalbaselineskip} 
\setlength{\textfloatsep}{0\normalbaselineskip}
\setlength{\abovecaptionskip}{0.cm}
\setlength{\belowcaptionskip}{-0.cm}
\centering 
\includegraphics[width=14.0cm, height=8.5cm, angle=0]{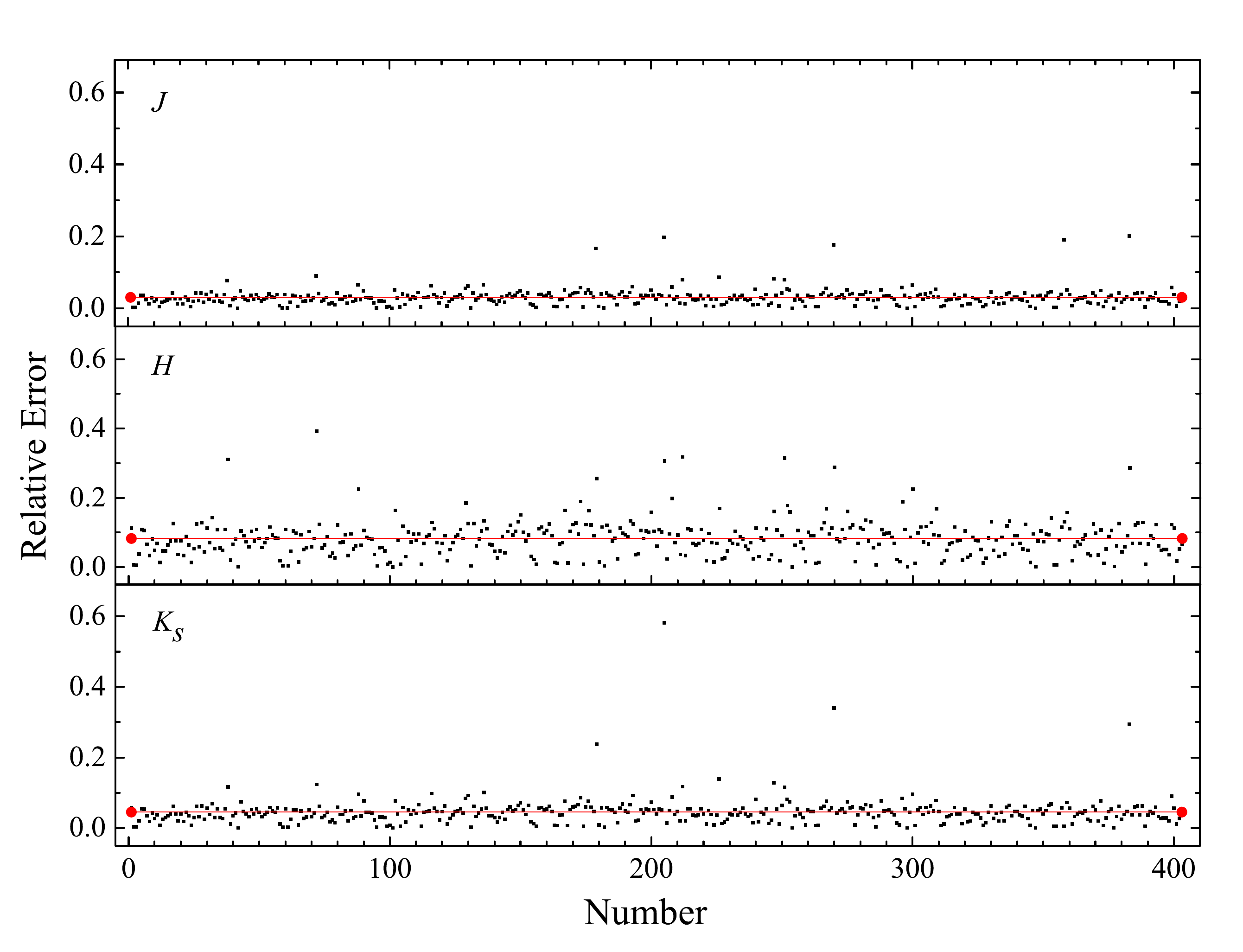}
\caption{ Relative error variation (2MASS).}
\label{fig:7}
\end{figure}

The specific variations of relative errors are presented in figure~\ref{fig:7} for these 3 bands. The errors for all bands are relatively concentrated and the variances are small, which can be used to invert the flux densities of the corresponding bands.

The SPSO algorithm is used to calculate the flux densities using the star catalogues, of which the MSX and 2MASS catalogues exhibit good inversion effects for all bands. For the LAMOST and WISE catalogues, the effect is not realized in some bands.

\begin{figure}[!htbp]
\setlength{\abovecaptionskip}{0.cm}
\setlength{\belowcaptionskip}{-0.cm}
    \centering
    \begin{tabular}{cc} 
        \includegraphics[width=7.0cm, height=5.1cm, angle=0]{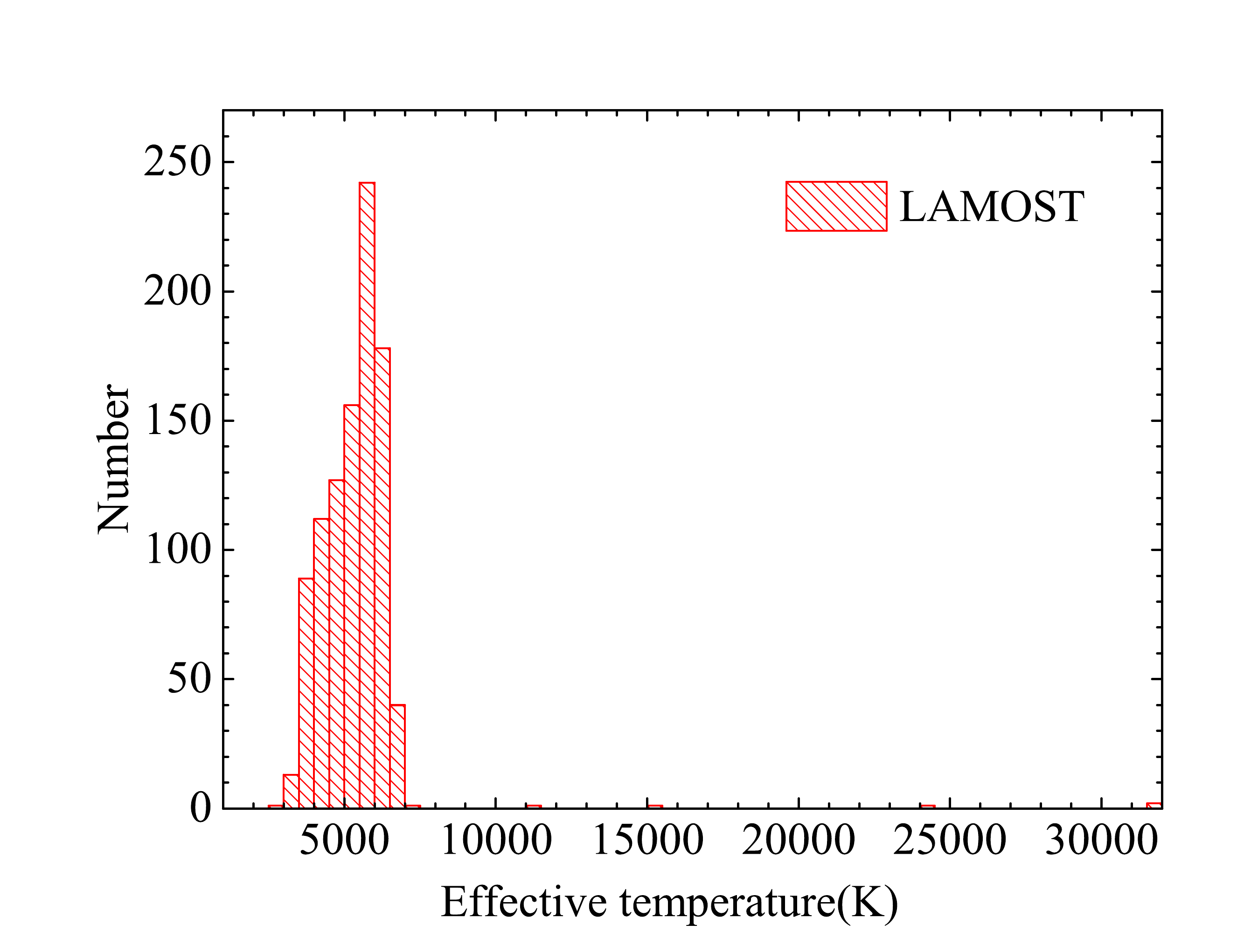} & \includegraphics[width=7.0cm, height=5.1cm, angle=0]{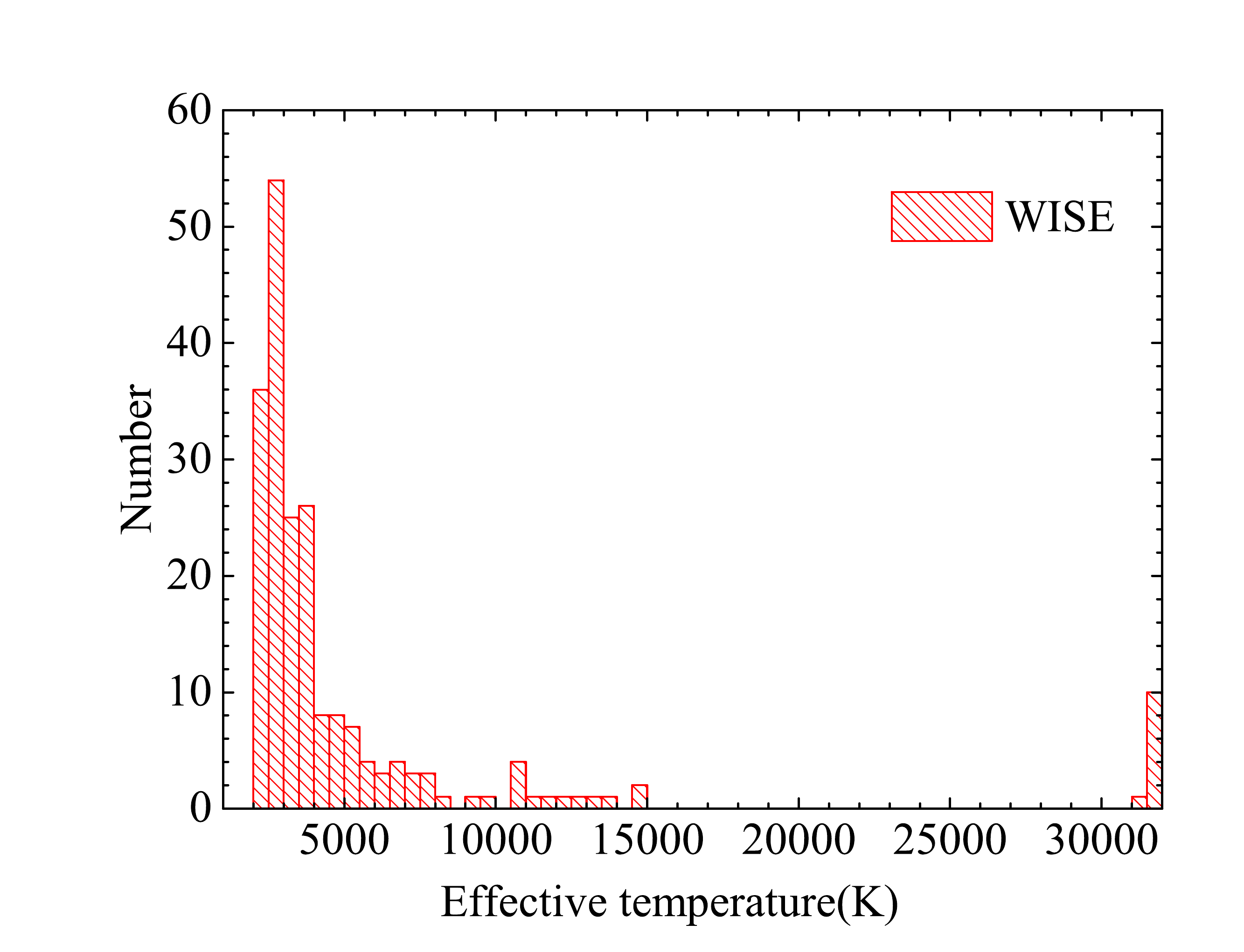}
        \\
        (a) LAMOST & (b) WISE
        \\
        \includegraphics[width=7.0cm, height=5.1cm, angle=0]{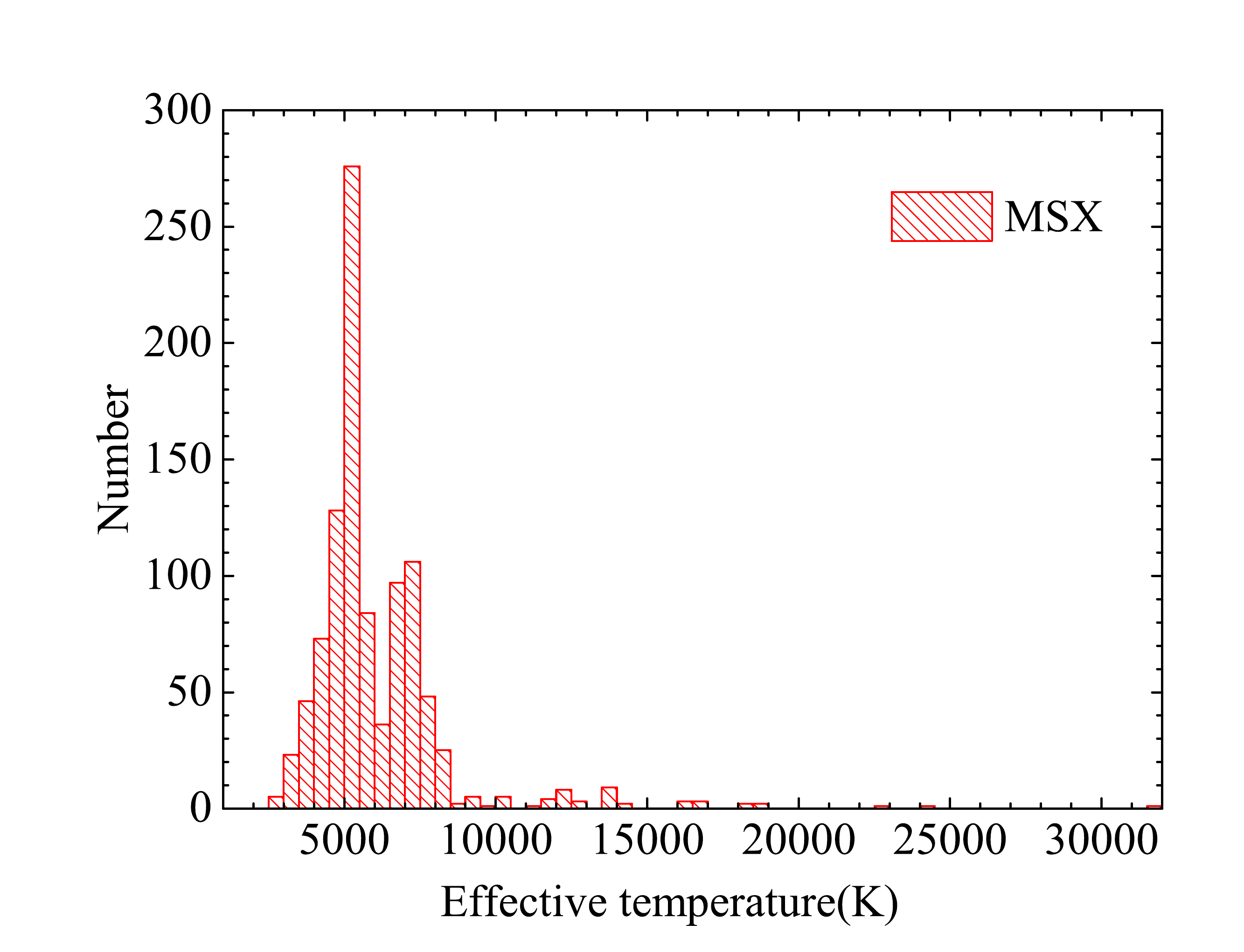} & \includegraphics[width=7.0cm, height=5.1cm, angle=0]{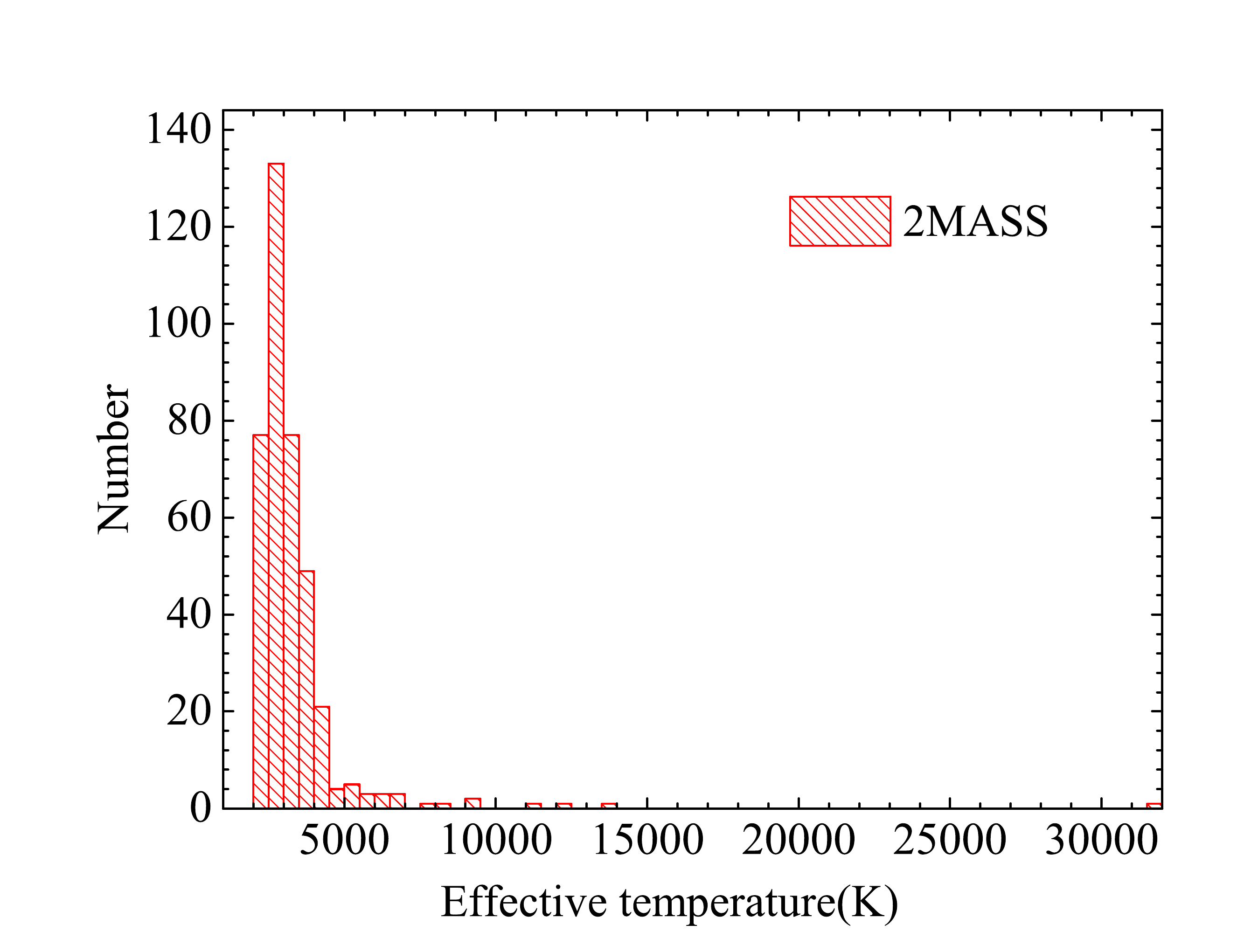}
        \\
        (c) MSX & (d) 2MASS
    \end{tabular}
    \caption{Histogram distribution of stellar effective temperatures in the four catalogues (LAMOST, WISE, MSX, 2MASS).}
    \label{fig:8}
\end{figure}

The histogram distributions of $T_{eff}$ in the four catalogues are plotted as figure~\ref{fig:8}. As can be seen, the stellar effective temperatures mainly range from 3000 K to 7000 K for LAMOST catalogue. As for WISE, the temperatures change from 2000 K to 8000 K. For MSX, the temperatures are 3000 K to 8500 K, and for 2MASS, 2000 K - 7000 K. The histogram distributions take 500 K as the interval. The maximum number of stars appears from 5500 K to 6000 K for LAMOST catalogue. For MSX, 5000 K - 5500 K. For both WISE and 2MASS, 2500 K - 3000 K. The temperatures of stars for MSX and LAMOST are higher than that for WISE and 2MASS.

\begin{figure}[!htbp]
\setlength{\abovecaptionskip}{0.cm}
\setlength{\belowcaptionskip}{-0.cm}
    \centering
    \begin{tabular}{cc} 
        \includegraphics[width=7.0cm, height=5.1cm, angle=0]{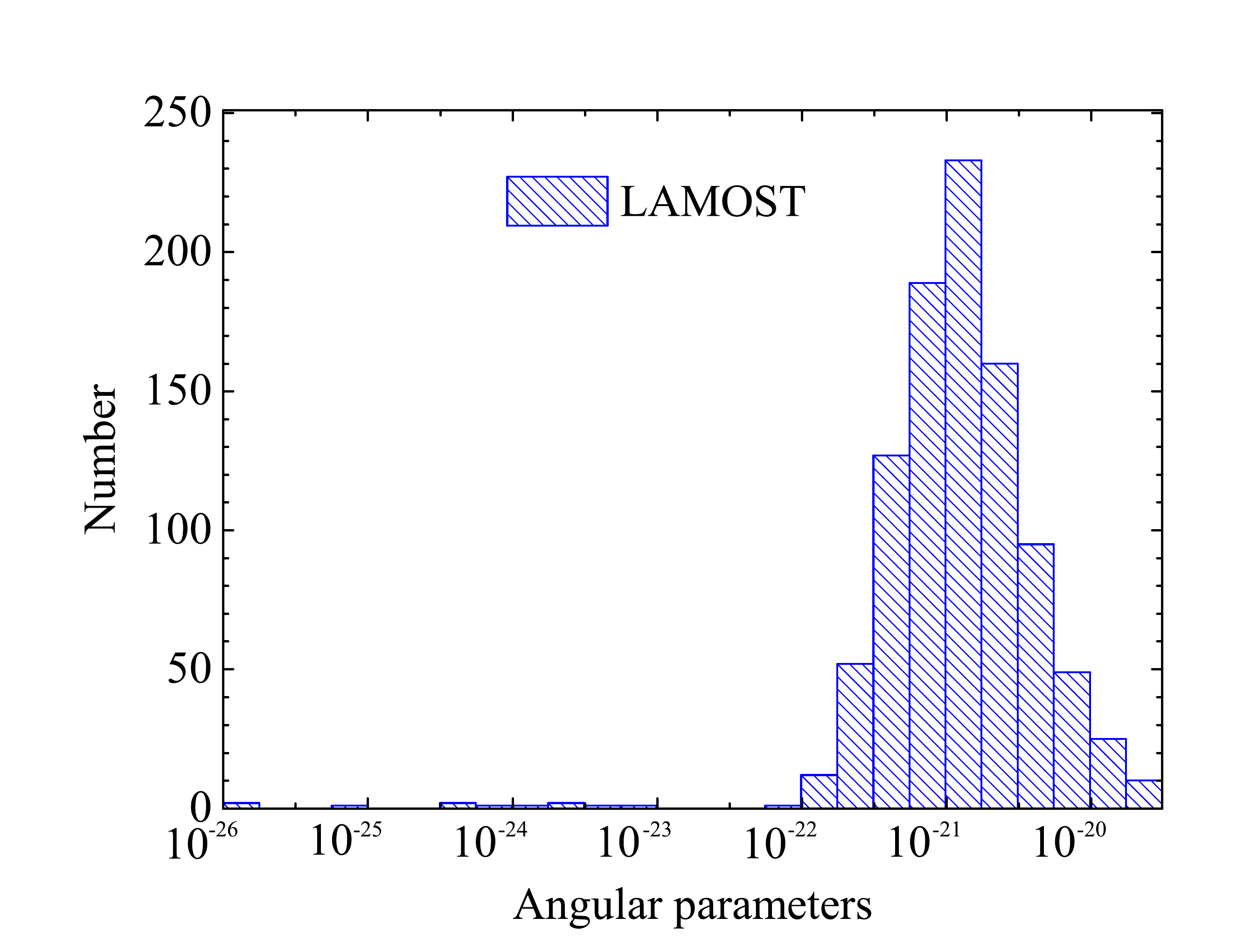} & \includegraphics[width=7.0cm, height=5.1cm, angle=0]{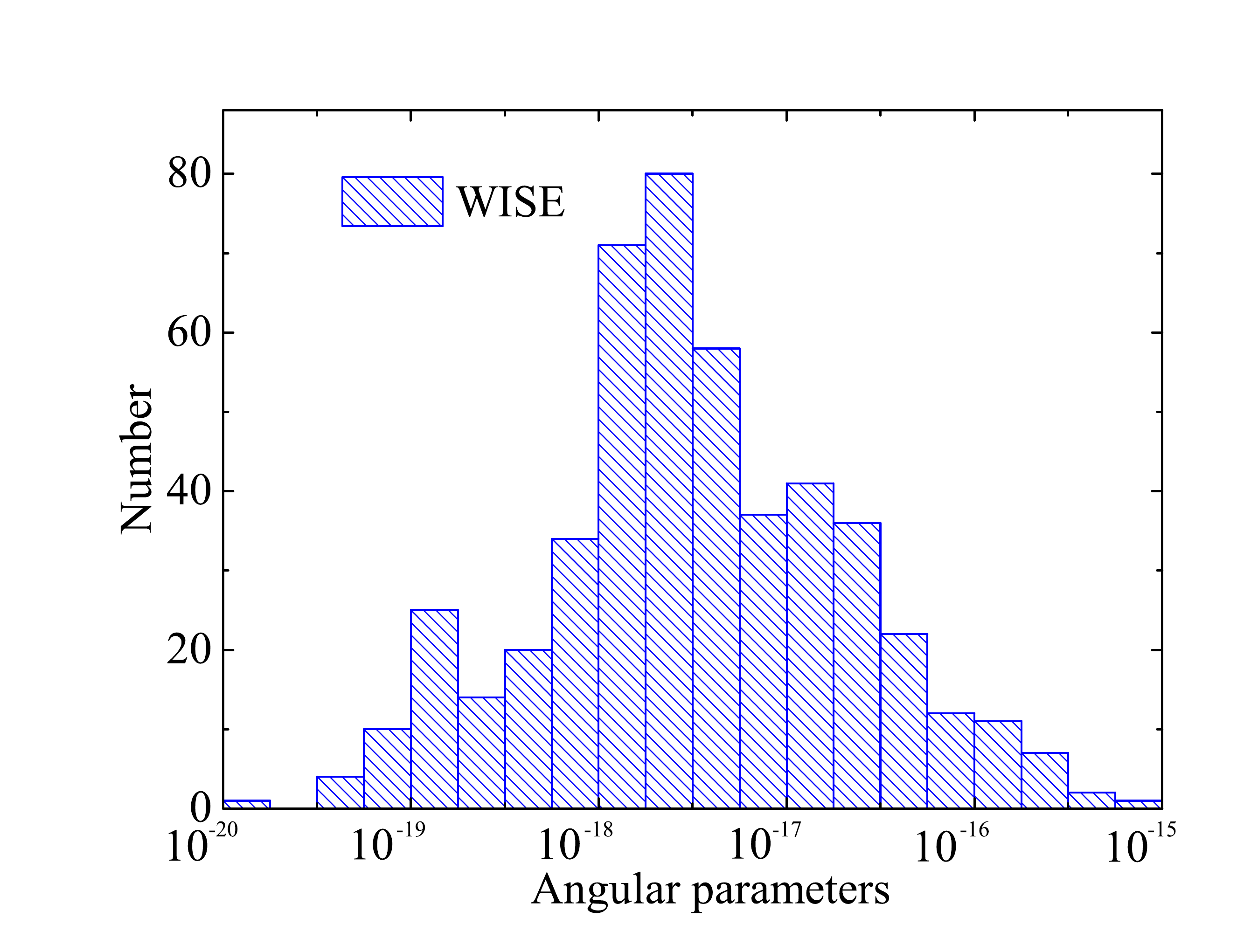}
        \\
        (a) LAMOST & (b) WISE
        \\
        \includegraphics[width=7.0cm, height=5.1cm, angle=0]{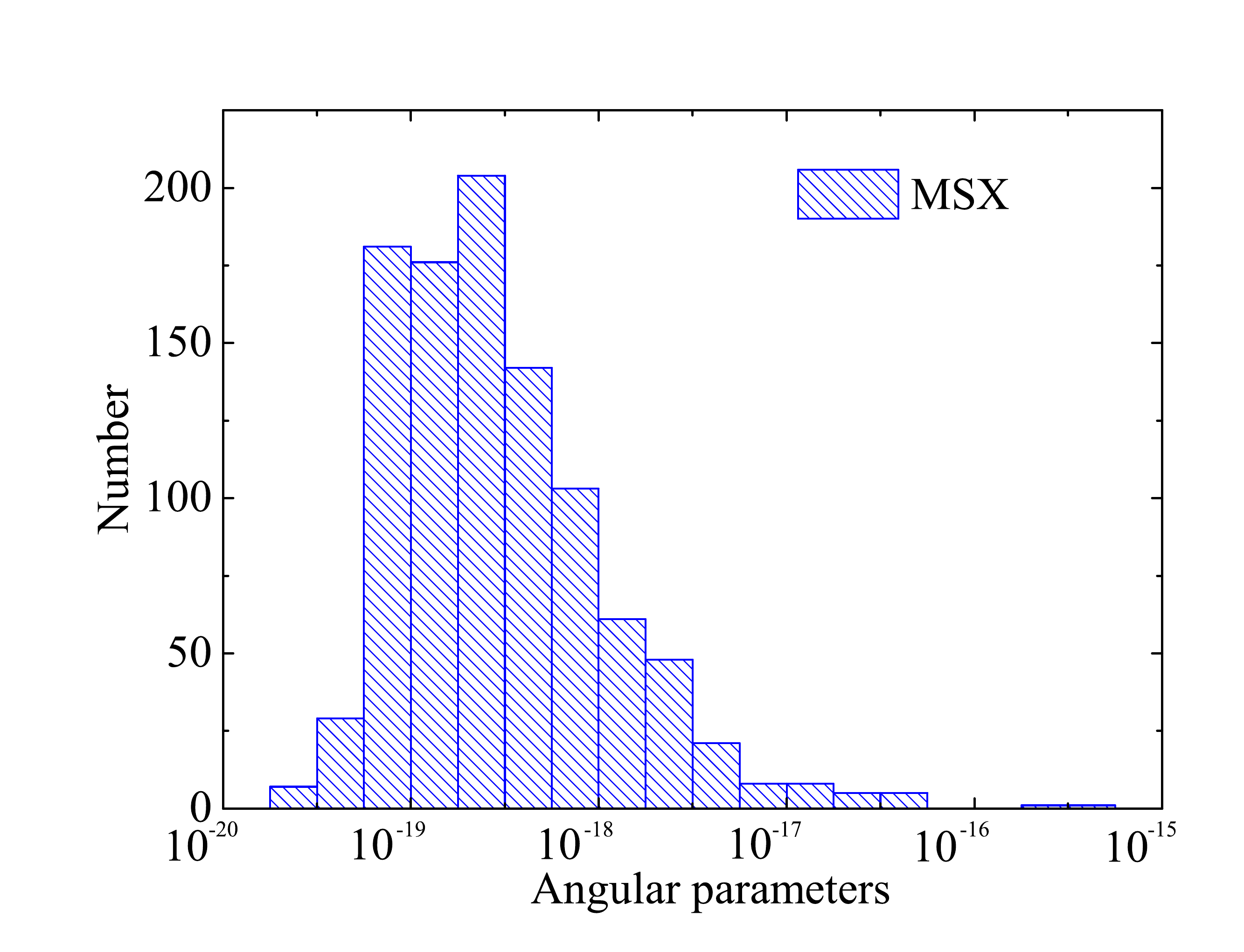} & \includegraphics[width=7.0cm, height=5.1cm, angle=0]{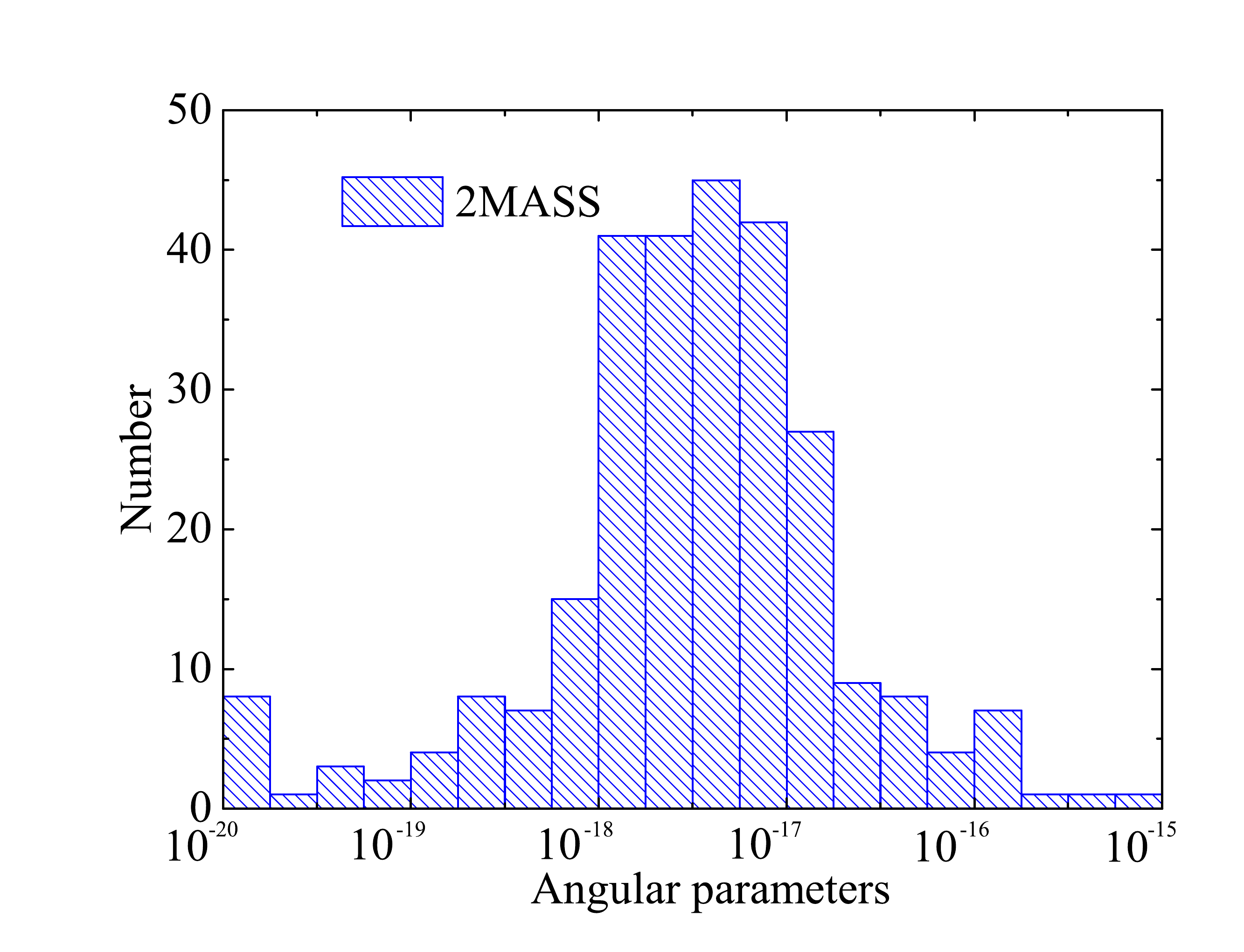}
        \\
        (c) MSX & (d) 2MASS
    \end{tabular}
    \caption{Histogram distribution of stellar angular parameters in the four catalogues (LAMOST, WISE, MSX, 2MASS).}
    \label{fig:9}
\end{figure}

The histogram distributions of stellar angular parameters in the four catalogues are plotted as figure~\ref{fig:9}. The angular parameters mainly change in the range 10$^{-26}$ - 10$^{-20}$ for LAMOST catalogue. As for other three catalogues, the angular parameters cover 10$^{-20}$ - 10$^{-15}$. For LAMOST catalogues the maximum number of stars is from 10$^{-21}$ - 10$^{-20}$. MSX: 10$^{-19}$ - 10$^{-18}$. WISE and 2MASS: 10$^{-18}$ - 10$^{-17}$. The angular parameters of stars for MSX and LAMOST are lower than that for WISE and 2MASS. The trend of angular parameters is opposite to that of temperatures. It verifies the correctness of the model.

Regarding the inversion results of the catalogues, the deviation of flux density for the selected band is smaller than that for the unselected bands. This condition can be explained as follows. The wavelength ranges and intervals are narrow for the bands of LAMOST catalogue, resulting in larger deviations in the results. The four central bands are 3.4, 4.6, 12, and 22 $\upmu$m for the WISE catalogue, and there is no intermediate transition band, resulting in larger deviations. There are no such deficiencies for the MSX and 2MASS catalogues, and the results have a high accuracy. Therefore, the flux densities of the given bands in the MSX and 2MASS catalogues can be calculated and provides data support for detection and identification.

\section{Conclusions}
\label{sec:conclusions}
Four star catalogues are studied and consist of the Large Sky Area Multi-Object Fibre Spectroscopic Telescope (LAMOST), Wide-field Infrared Survey Explorer (WISE), Midcourse Space Experiment (MSX), and Two Micron All Sky Survey (2MASS). The stellar flux densities in the given band range are obtained using the SPSO algorithm combined with the radiative energy of the known bands of star catalogues. The particle number of the algorithm is optimized in the inversion calculation. The flux densities of the LAMOST, WISE, MSX, and 2MASS star catalogues are calculated. The following conclusions are obtained.

1. The accuracy of the flux densities for different band combinations is different for the star catalogue data. The best band combinations are C1, A2, and C3 for the LAMOST, WISE, and MSX catalogues, respectively. They have lowest relative errors and satisfy the requirement of selecting the best band combination. The relative errors in some bands are large for the LAMOST and WISE catalogues. The energy data of the 2MASS catalogue can be used as an input source in the inversion calculation of flux densities of given bands.

2. Comparing the energy data of four star catalogues, it is found that the flux densities of the inversion calculation are in good agreement, the variance is small, and the data are more concentrated for the MSX and 2MASS catalogues. The flux densities in some bands are very poor, the variance is large, and the data are highly scattered for the LAMOST and WISE catalogues. They need to be modified in these bands. Therefore, the energy data of the MSX and 2MASS catalogues can be used as an input source to calculate the flux densities of the given bands in this model. They can provide data support for detection and identification.

\section{Acknowledgements}
This work was supported by the National Natural Science Foundation of China (Grant Nos. 51776051 and 51406041). We would like to thank the editors and referees for their comments, which helped us to improve this paper.


\end{document}